# Optical and Electronic Properties of Colloidal CdSe Quantum Rings

[1]James Xiao, [1]Yun Liu, [2]Violette Steinmetz, [1]Mustafa Çağlar, [1]Jeffrey Mc Hugh, [1]Tomi Baikie, [1]Nicolas Gauriot, [1]Malgorzata Nguyen, [1]Edoardo Ruggeri, [1]Zahra Andaji-Garmaroudi, [1,3]Samuel D. Stranks, [2]Laurent Legrand, [2]Thierry Barisien, [1]Richard H. Friend, [1]Neil C. Greenham, [1]Akshay Rao* and [1]Raj Pandya*

* correspondence: ar525@cam.ac.uk, rp558@cam.ac.uk

[1]Cavendish Laboratory, University of Cambridge, J.J. Thomson Avenue, CB3 0HE, Cambridge, United Kingdom

[2]Sorbonne Université CNRS-UMR 7588, Institut des NanoSciences de Paris, INSP, 4 place Jussieu, F-75005 Paris, France

[3]Department of Chemical Engineering & Biotechnology, University of Cambridge, Philippa Fawcett Drive, CB3 0AS, Cambridge, United Kingdom

Abstract

Luminescent colloidal CdSe nanorings are a recently developed type of semiconductor structure that have attracted interest due to the potential for rich physics arising from their non-trivial toroidal shape. However, the exciton properties and dynamics of these materials with complex topology are not yet well understood. Here, we use a combination of femtosecond vibrational spectroscopy, temperature-resolved photoluminescence (PL), and single particle measurements to study these materials. We find that on transformation of CdSe nanoplatelets to nanorings, by perforating the center of platelets, the emission lifetime decreases and the emission spectrum broadens due to ensemble variations in the ring size and thickness. The reduced PL quantum yield of nanorings (~10%) compared to platelets (~30%) is attributed to an enhanced coupling between: (i) excitons and CdSe LO-phonons at 200 cm$^{-1}$ and (ii) negatively charged selenium-rich traps which give nanorings a high surface charge (~-50 mV). Population of these weakly emissive trap sites dominates the emission properties with an increased trap emission at low temperatures relative to excitonic emission. Our results provide a detailed picture of the nature of excitons in nanorings and the influence of phonons and surface charge in explaining the broad shape of the PL spectrum and the origin of PL quantum yield losses. Furthermore, they suggest that the excitonic properties of nanorings are not solely a consequence of the toroidal shape but are also a result of traps introduced by puncturing the platelet center.

The shape and size of a semiconducting nanocrystal is key to determining both its physical and optoelectronic properties *via* the nature of exciton quantum confinement.[1] Where spherical nanocrystals have 0D confinement, rod- and plate-shaped nanocrystals have 1D and 2D confinements leading to, for example, linearly polarized emission which is free of inhomogeneous broadening.[2] Over the last two decades, a great amount of knowledge has been gained in synthesising these nanocrystals and their heterostructured variants (*e.g.* dot-in-plate),[3] with cadmium chalcogenide (CdX, X = S, Se, Te) materials finding particular use in a range of fields from catalysis to neuronal voltage sensing.[4-,6]

Recently the shape of CdSe nanocrystals has been extended to nanorings by Fedin *et al*.[7] Where dots, rods, *etc*. all belong to the topological class with genus (g) equal to zero (g = 0) and Euler characteristic $\chi = 2$, rings have g = 1.[8] Due to synthetic challenges ring-like topologies are challenging to prepare. Epitaxial semiconductor rings have been shown to exhibit a range of unusual properties including terahertz absorption and the formation of magnetoexcitonic quantum states (Aharonov-Bohm effect).[9,10] Coupling these phenomena with a luminescent solution processable material would be highly desirable for a range of magneto-optical applications *e.g.* data storage or Faraday rotators.[11,12] Indeed, although there are currently a large range of shapes available for nanostructures: dots, rods, plates, tetrapods, ribbons, *etc*., the ring geometry sits at the interface between 0D and 1D confinement which is stronger than in dots, but likely weaker than in platelets.[13] Combining aspects of both, *e.g.* narrow linewidths and polarisation properties, could prove to be useful for display and lighting applications. In terms of device performance, the shape of the nanorings could result in highly polarized emission.[14,15] This would allow for an increase in the efficiency of LEDs by removing external polarising optics used to minimise the reflection of ambient light. Additionally, highly directional emission *via* the alignment of dipoles in the ring plane could remove the need for external focussing optics a common issue in practical LEDs. Highly directional emission from this geometry would also be useful in luminescent solar concentrators.[16] Nanoring geometries have also been shown to be promising for high harmonic generation[17] and the potentially large absorption cross section and transition dipole moments of nanorings, combined with their polarised emission, could be used for applications in strong coupling.[18] The nanoring geometry has been used in photovoltaic cells as a way to stack nanoparticles without the need for strain compensation, allowing for potentially higher power conversion efficiencies .[19] Colloidal nanorings could hence be useful in quantum dot sensitized solar cells.[20] Many metamaterials are also based on nanoring geometries.[21] Combining the semiconducting properties of CdSe nanorings with appropriate patterning might allow for hitherto unexplored devices with additional features such as negative refraction.[22] One recent avenue that has been explored with nanocrystals is as neuronal voltage sensors due to their high photoluminescence stability and response to electric fields by the quantum confined Stark effect.[6,23] However, typically this has only been explored in the realm of 0D nanocrystals, whose size is comparable to the thickness of the plasma membrane (~5 – 10 nm), making them ineffective local probes. Using nanorings however, which have a thickness of ~1 – 2 nm and can localise effectively in different parts of the membrane could improve the viability of such quantum dot voltage sensors. The exposed facets in terms of the ring hole offers a prospect for chemistry and catalysis not achievable with other nanoparticle shapes.[24] Indeed, threading, isolating or trapping molecules in the centre of rings is an attractive target for supramolecular colloidal science.[25] This is in addition to the aforementioned fundamental physics playground this geometry offers. The potential for nanorings to support both Rashba-type and spin-orbit type interactions also is appealing for use of these materials in dual optoelectronic and spintronic applications.[26]

Despite this, the electronic and optical properties of colloidal CdSe nanorings has remained largely unexplored, and although work by Hartmann *et al.* has shown that excitons in rings possess an unusual arrangement of in-plane linear dipoles, the exciton dynamics have not been established.[15] Here, we use a combination of ultrafast transient absorption spectroscopy and temperature-resolved optical and structural measurements to achieve an in-depth characterization of CdSe nanorings. Using absorption and TA measurements we establish that even a partial etch of the platelet induces a red-shift, a broadening of the absorption spectrum, and a decrease in the excited state lifetime. Density functional theory (DFT) calculations show that this is a result of changes in the electronic structure and overall material thickness in the direction of quantum confinement introduced by etching. The emission lifetime also decreases as the platelet is etched to form a ring, likely arising from the activation of phonon and trap mediated non-radiative decay pathways. Single particle PL measurements show the solution PL is inhomogenously broadened, likely arising from different sizes in the central hole from ring to ring, leading to non-uniform decay of the PL across the emission band. The emission quantum yield of the rings (~10%) is markedly lower than platelets (~30%), leading to poor energy transfer between nanorings. We find this to be a consequence of coupling between excitons, CdSe LO-phonons at 200 cm$^{-1}$ and negative selenium rich traps. We quantify these charge effects with environmental Kelvin probe and zeta potential measurements. The structural modifications induced by etching of the platelet center systematically increase exciton-phonon coupling even when the platelet is only partially etched. Additionally, we find population of sub-band gap trap like states also governs the temperature resolved emission of rings. We quantify the energy barrier for escaping from these traps to be ~8 meV, markedly lower than equivalent states in platelets (~21 meV). Although the temperature dependent PL behaviour of nanoplatelets and nanorings are similar, we use time-resolved measurements to demonstrate the electronic structure cannot simply be treated as a bright-dark doublet in the latter.[27] Finally, we qualitatively find that Auger and carrier cooling effects are less significant in rings than in platelets, with the excited state lifetime of the former independent of both excitation energy and intensity. We ascribe this to a hopping-type picture where excitations sit in sub-bandgap trap sites on the nanoring surface as compared to the delocalized band like picture of excitations in platelets. Using a variety of models, we interpret the coupling between exciton (core) and surface state in these materials and quantify the number of trap sites per particle. Finally, we also suggest rational protocols in terms of ligands, particle sizes and aspect ratios for reproducible synthesis of this relatively recently explored colloidal nanocrystal geometry. Our results provide a detailed description of excitons in a ring-like luminescent semiconductor, whose colloidal synthesis means its structure is drastically different from the "volcano-like" structures obtained using epitaxy.[28,29]

Our findings will be of use not only for realizing the potential of colloidal nanorings, but more generally in designing topological semiconductors for optoelectronic and quantum technologies. The wet chemical synthesis methods further explored in this work can be adapted to the wide family of II-VI semiconductor materials with similar chemistries spanning a broad spectral range. By understanding how defects deviate the properties of these materials away from ideality we will be able to better engineer the PLQY, lifetime *etc.*, for such applications. Indeed, a wide range of techniques, often not (simultaneously) used in the study of colloidal nanostructures, *e.g.* impulsive vibrational spectroscopy, Kelvin probe *etc.*, are explored here. The work demonstrates the power of these methods to achieve insights into trap states in nanocrystals, and how these techniques might be generally applied to other nanocrystalline systems. We present these less explored techniques as useful additions to the current suite of characterisation techniques and analyses commonly employed, with particular emphasis on shape-controlled nanocrystal systems, which expose atypical surfaces and trapping sites. The application of trapping models originally developed for 0D dots to the plate and annular geometries demonstrates the versatility of these methods regardless of nanocrystal shape. Studying the transformation of nanoplatelets into nanorings in a *quasi-in-situ* manner provides correlated structure-function insights, *e.g.* between etch-time and exciton-exciton annihilation, bringing us closer to the real-time tracking of the electronic structure in nanomaterials as they form. We also identify requirements for the synthesis of more complex topologies with multiple fused rings, and although this work is focussed on a single ring system, a basis for studying more intricate systems has been established for future work. In these structures the quantum confinement should be stronger than dots or single rings.[13] This allows for single bound states which may be suitable for terahertz intersublevel detectors. Although many of the aforementioned exotic properties of nanorings are not yet realised in this work, the fundamental insights into the electronic structure gained permits rational optimization and application of these materials in devices for quantum technologies, optoelectronics and spintronics.

Results and discussion

CdSe nanorings as shown in Figure 1a were synthesized using a method previously reported by Fedin *et al.*[7] Briefly, 4-monolayer thick CdSe nanoplatelets were first synthesized , and showed an excitonic emission at 512 nm.[2,30] A solution of these nanoplatelets was then heated with Se powder for 5 – 10 min to etch the platelet center followed by quenching of the reaction with tributylphosphine. Nanoplatelets with a slightly irregular shape (Figure 1a, left) compared to square or rectangular nanoplatelets which are often routinely synthesised were found to give the 'cleanest' ring etch and most monodisperse samples as evidenced from TEM. [22,30] A maximum nanoplatelet aspect ratio of 1.5:1 (length:width) and maximum dimension of ~15 nm were also found to be important for the etch to give nanorings with a single hole. Further details of sample preparation can be found in the Methods section. Unless otherwise stated all measurements were carried out in solution on rings with a hole dimension of ~5 – 10 nm. Samples were always measured within 3 days of preparation and stored and measured in an oxygen-free argon environment, to minimize oxidative degradation.

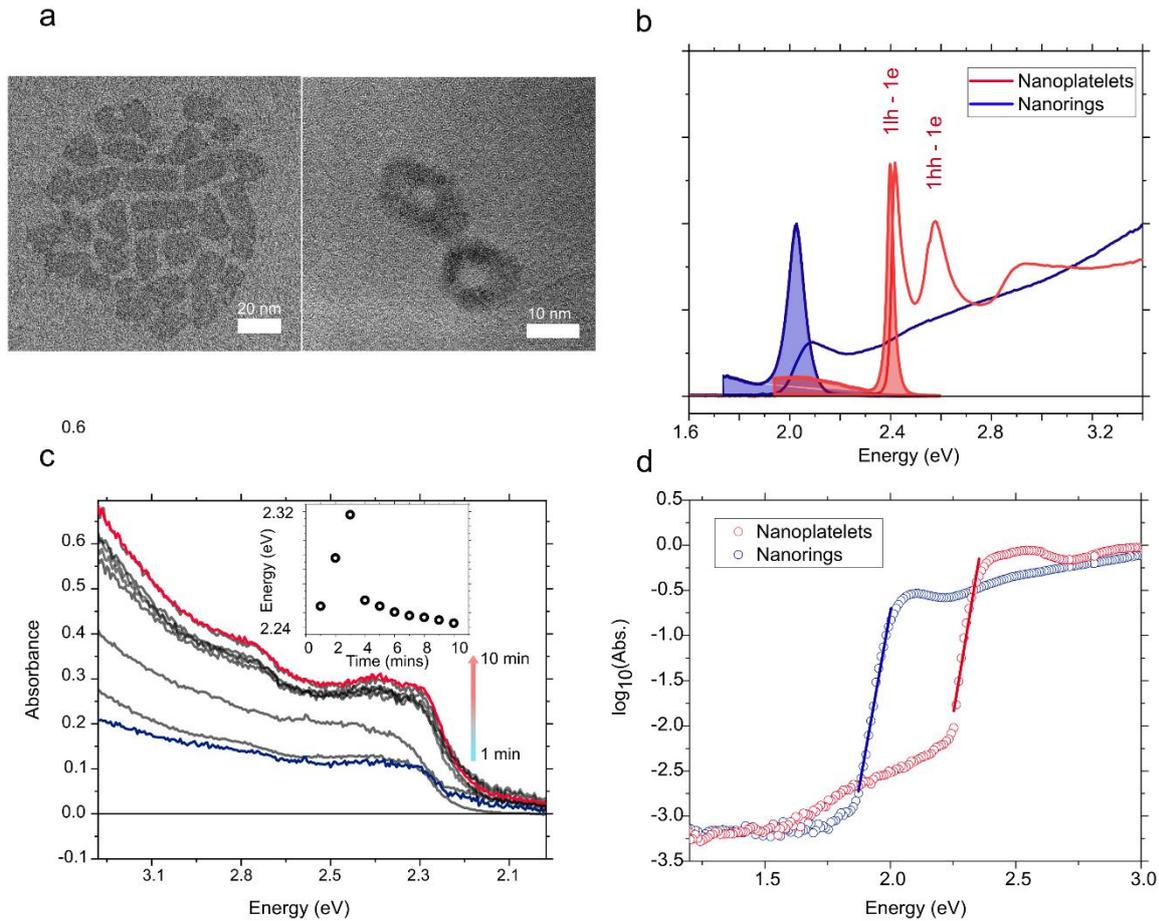

**Figure 1: Structural and optical characterization of CdSe nanorings. a.** Transmission electron microscopy (TEM) image of CdSe nanoplatelets (left) and nanorings (right). Nanorings are formed following etching of the nanoplatelet center. The scale bars in the two images are 20 nm and 10 nm respectively. **b.** Absorption spectrum of CdSe nanoplatelets (blue) and nanorings (red) in solution. Nanoplatelets show excitonic peaks at 2.57 eV and 2.4 eV corresponding to the electron-heavy hole and electron-light hole transitions. In nanorings there is a broad excitonic absorption centered at 2.1 eV. The respective emission spectrum for 405 nm laser excitation is shown in shading[33]. **c.** Absorption spectrum of nanorings as a function of etch time. The nanoplatelet excitonic peaks mostly disappear within the first minute of etching and are replaced by a broad absorption peak. The nanoring absorption edge (inset) red shifts over the etch time (10 mins) with the absorption growing in intensity. The hh-1e or lh-1e transition in nanoplatelets potentially appear as a weak shoulder in the absorption spectra of the rings at ~2.8 eV. TEM images (**SI, S1**) show that in 1 – 6 min the majority of nanorings remain partially etched (colour of spectra match arrow. **d.** Photothermal deflection spectrum (PDS) of CdSe nanoplatelets (red; ~50 nm thick film) and nanorings (blue; ~75 nm thick film). Fitting the absorption tail, we find an Urbach energy of ~40 meV for nanoplatelets and ~38 meV for nanorings. In nanoplatelets there is a broad tail of absorbing states out to 1.5 eV these have been assigned to defects and inter-sub-band transitions.[34]

To understand better the transformation of nanoplatelets to nanorings *in-situ* absorption measurements were performed. Here, aliquots of solution were taken at given times during etching. Figure 1b, c shows the absorption spectra for 4-monolayer (ML) nanoplatelets and nanorings in solution. In order to minimize the effects of scattering, measurements were carried out within an integrating sphere (see Methods). In contrast to the sharp absorption and weakly Stokes shifted emission of nanoplatelets, nanorings show a broad absorption with an emission that is red-shifted from the absorption peak. Immediately (within 1 min) after reaching reaction temperature it can be seen that the sharp CdSe 1lh-1e and 1hh-1e excitonic transitions[30] mostly disappear and are replaced with a broad absorption feature that gradually red shifts as a function of etch time (although we note that the predominant optical changes take place below our time resolution, within 1 min of etching). This additional red-shift is most likely a consequence of the increased thickness of the rings over that of the initial platelets. This will in turn alter the nature of quantum confinement and mitigate any blue-shift that would be expected from the extra confinement conditions introduced by having excitations on a ring. An additional absorption shoulder around 500 nm also grows in throughout the etching process. This peak may correspond to the lh-1e or hh-1e transitions observed in nanoplatelets, which in nanorings are mixed with the localized valence band maximum state (see DFT below). The poor spectral resolution however precludes further discussion. Because there is a range of ring thicknesses (along the long and short-axes of the ring) and variation in the inner and outer ring diameters, following addition of the etchant, the absorption peak is broad and more closely resembles that observed in 0D dots.[1] Structural measurements performed in the initial work of Fedin *et al.* additionally confirm increase in thickness on transforming platelets to rings.[15] We note that the TEM image of nanorings in Figure 1a show some variation in contrast suggesting the thickness of individual rings is non-uniform. Although this will increase the electronic disorder of nanorings, it does not appear to lead to multiple *emissive* excitonic sites as demonstrated by single particle PL measurements (see later).

To understand this change in the absorption spectra further, we performed DFT calculations. We used a nanoplatelet infinite in lateral dimensions and 2 CdSe monolayers thick to represent the pseudo-2D nanoplatelet synthesized in experiments, which possess lateral dimensions much larger than the Bohr exciton radius (~5.4 nm).[35] The top and bottom {100} surfaces of the platelet were passivated with chloride ligands to ensure overall stoichiometry.[36,37] The nanorings were created by removing atoms from the platelets to form holes of different sizes. Further details about the DFT calculations can be found in the Methods. As seen in Figure 2a, the platelet exhibits the two absorption peaks due to the 1lh-1e and 1hh-1e. These peaks are red-shifted compared to experiments due to the well-known underestimation of the semiconductor bandgap by DFT. Compared to the platelet, the first absorption peaks of the nanorings are red shifted by about 150-300 meV, in qualitative agreement with experiments. Although we note this peak has a relatively low intensity and because exciton effects are not included, our calculations likely only gives a qualitative picture. As indicated by Figure 2b, when a small ring is initially formed, a set of localized states (Figure 2c) emerge just above the valence band maximum (VBM). These states are mainly of Cl characteristics (**SI, S2i**) and are localized on opposite sides of the ring. The wavefunction localization pattern is consistent with the *k*-vector resolved PL emission reported by Hartmann *et al.,* whereby the measured radiation pattern shows bright lobes on opposite ends of the ring.[15] The change in topology only affects the VB states, and the conduction band states near the Fermi level remain delocalized and energetically similar to those of the nanoplatelet (**SI, S2ii**). We do note that our calculation is for the stoichiometric nanoring structure whereby all dangling bonds were passivated, but in experiments the existence of dangling bonds could introduce additional trap states and changes in the electronic structure.[32,38]

Our DFT results suggest that the energy shifts in the experimental absorption spectra likely arises from two effects: (i) the redshift in the absorption peak as more materials are pushed onto the rings as the etching proceeds, increasing the thickness in the quantum confinement direction and (ii) the blueshift in the absorption peak as the size of the ring increases.

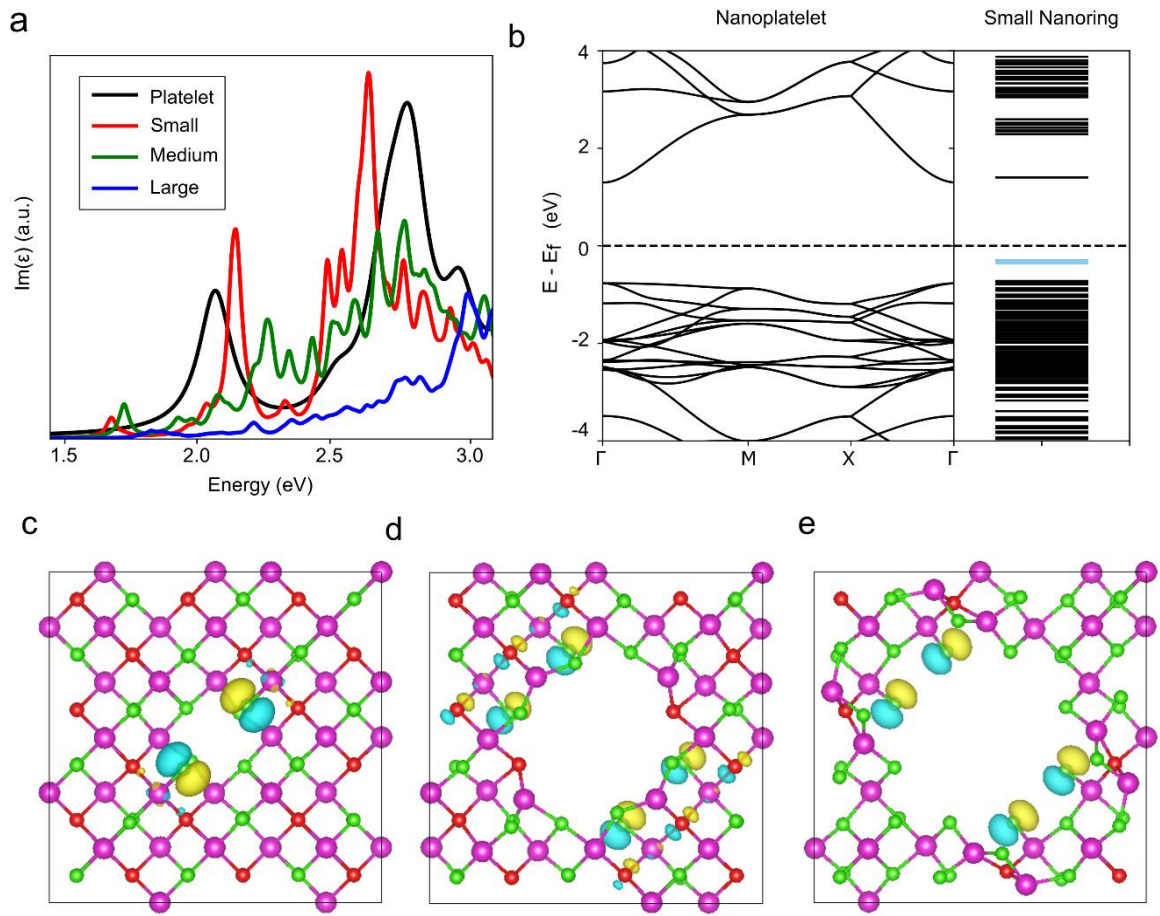

**Figure 2: DFT calculations of nanoplatelet and nanoring. a.** In-plane absorption spectra (imaginary part of the dielectric function) of the nanoplatelet compared with nanorings of different sizes. **b.** The band structure of the CdSe nanoplatelet, with the Fermi level denoted by the dashed line. The special *k*-points in the Brillouin zone of the reciprocal lattice are $\Gamma = \{0,0,0\}$, $M = \{0.5, 0.5, 0.5\}$ and $X = \{0.5, 0.0, 0.0\}$. The band energies at the $\Gamma$ point of the small nanoring compared to the nanoplatelet. The localized states at the top of the VBM are in blue. **c- e.** The in-plane view of the ball-and-stick model of the nanorings of small, medium and large sizes respectively. The Cd, Se and Cl atoms are represented by magenta, red and green spheres. The isosurfaces of the VBM wavefunction are colored cyan and yellow. The data are visualized using VESTA.[39]

To understand the role of disorder introduced by etching of the platelet center were performed using photothermal deflection spectroscopy (PDS), which is insensitive to scattering (Figure 1d; Methods).[40] The PDS absorption spectra of rings shows a single excitonic peak at ~2 eV followed by a relatively sharp drop. For platelets a similarly sharp drop in the absorption is observed around 2.25 eV but this levels to a weak and broad absorption feature around 2 eV likely related to the absorption of sub-gap or defect states.[34,41,42] Having identified the relevant tail transitions, we can use the PDS measurements to quantify the energetic disorder of both rings and nanoplatelets. This is characterized by the Urbach energy, $E_u$, which is related to the absorbance of the material, $A$, by $A(E) \propto e^{\frac{E}{E_u}}$, where $E$ is the photon energy.[43] Fitting the absorption tail in Figure 1d to the Urbach formula gives an $E_u$ value of $40 \pm 1$ meV for CdSe platelets and $38 \pm 1$ meV for CdSe nanorings. This value is approximately 4 times larger than that obtained for high-quality GaAs, but lower than values obtained for CdS and CdSe/CdS nanocrystals ($E_u \approx 48$–65 meV) and common organic semiconductors (P3HT $\approx 50$ meV).[33,44-46] This suggests that although there is a change in the electronic structure on etching of nanoplatelets (with potential thickness variations within a single ring) these changes do not significantly impact the degree of electronic disorder. This is likely because the surfaces/edges of the nanostructures, common to both, is where most of the traps contributing to the Urbach energy are located. We note in nanoplatelets there is an additional tail of absorbing states between 1.5 – 2 eV, which may arise from surface defects or sub-band gap states.

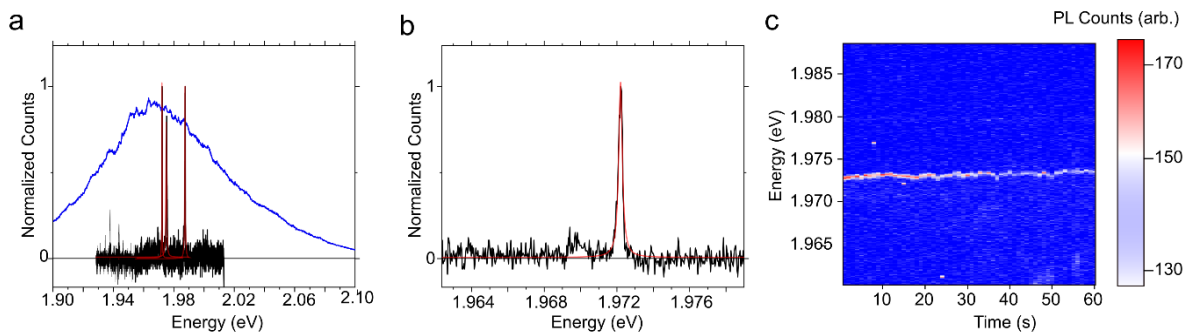

**Figure 3: Photoluminescence spectra of single CdSe nanorings at 4 K. a.** Typical emission spectra of single CdSe nanorings (black lines; each of the spectra is taken from a *different* single object). Red overlay shows a Lorentzian fit to the spectra, with blue spectrum indicating photoluminescence from ensemble. A range of emission energies are observed for single nanorings suggesting the ensemble spectrum to be composed of rings of varying thickness and/or diameter. Both ensemble and single particle spectra are measured at 4 K **b.** Zooming into an example single nanoring PL spectrum shows a line width of ~250 μeV in FWHM. **c.** Time-dependent spectrum of emission from a single nanoring at 4K. The emission intensity decreases as a result of photobleaching (**SI, S3**), with no new peaks growing in and relatively little peak shifting.

The photoluminescence spectrum of nanorings in solution is relatively broad (FWHM ~90 meV) compared to nanoplatelets (FWHM ~ 35 meV). To understand whether this originates from sub-structure within the emission spectrum or inhomogeneous broadening effects, low-temperature (4 K) single-particle PL measurements were performed. There is no phase change in the structure of either nanoplatelets or nanorings on cooling to 4 K (**SI, S3**). As exemplified in Figure 3a-b single nanorings show a single sharp emission peak varying in center frequency between 1.94 and 2.00 eV and 250 – 700 μeV FWHM in width. This behavior is similar to nanoplatelets where there is large difference in linewidth between the ensemble (30 meV) and single particles (400 μeV) at 20 K.[47]

If exciton-phonon coupling was the dominant PL broadening mechanism in nanorings as is the case for CdSe/CdTe nanoplatelets for instance, one would expect this to manifest itself also in the single particle spectra *e.g.* through phonon sidebands.[48] Similarly, if emission from additional states, *e.g.* trions, made up the ensemble PL response we might expect these to manifest themselves in the single particle spectra also or as spectral jumps in Figure 4c. This latter figure shows the single nanoring PL peak shifts >30 μeV over 60 s, as a consequence of spectral diffusion, suggesting that multiple emissive states are not present. Ruling out homogeneous (lifetime) broadening is challenging without access to the single-particle radiative lifetimes (ensemble lifetimes are on the order of nanoseconds at 4 K), but we note that on increasing the excitation power the single-ring PL spectrum remains unchanged (albeit with rapid photobleaching) suggesting exciton-exciton interactions do not contribute significantly either. Finally, repeating single-particle PL measurements over ~14 nanorings (**SI, S3**) reproduces well the low-temperature ensemble spectrum. All together this suggests that inhomogeneous PL broadening dominates in nanorings, with the heterogeneity in the single-particle spectra explaining the broad (room temperature) ensemble linewidth. This shows promise for the use of colloidal nanorings in quantum optical applications which require coherent manipulation of states. Furthermore, it suggests that despite thickness variations within a ring confinement remains predominantly 1D, and that the main quantization axis is either related to the hole size as this is invariant within the ring or is just dominated by the particle thickness. The ensemble PL properties might be greatly improved by future synthetic methods such as automated microfluidic flow synthesis.[49] We note that the low PLQY (~10 %), ~90 meV Stokes shift and lack of stacking, also suggests significant FRET (see further discussion later) does not occur between the nanorings and they act as independent emitters. We also comment that the PLQY of both nanoplatelets and nanorings does not vary with excitation energy suggesting that hot carrier effects are marginal (see SI, S10) [50].

Based on our DFT calculations nanorings with a larger radius are expected to have a larger bandgap and contribute to the higher energy emission within the ensemble, with smaller rings showing red-shifted emission, potentially explaining the observed variation in emission energies. Alternatively, the distribution in nanoring bandgaps may arise from variations in thickness between rings induced during the etching process, with thicker rings having more red-shifted emission. We note that the observation of a single emission peak in the single-particle spectra suggests any variation in thickness still only leads to a single emissive site. In any case, the narrow single-particle line widths suggest that nanorings have potential for optical applications such as single photon sources, where the emission can be coupled with the unusual radiation pattern (Hartmann *et al.* showed the transition dipole moments in CdSe nanorings are arranged in a uniaxial, in-plane manner) arising from the annular shape.[15] It is challenging to quantitatively determine the distribution in ring sizes in the ensemble, however dynamic light scattering (**SI, S4**; DLS) measurements can be used to qualitatively estimate this. Here we find an average (total) nanoring size (hydrodynamic radius) of 15 ± 4 nm, with the range of the size distribution obtained from DLS being 28 nm (taken as 2σ from the mean size; [8 – 36 nm]). For nanoplatelets the distribution range is smaller at 13 nm [15 – 28 nm]. In both cases a spherical particle model is used to extract particle distributions and the results should be treated only qualitatively. However, the relatively large variation in ring sizes (compared with nanoplatelets) combined with any variations in the confinement in the transverse direction of the rings (thickness) likely contributes to the large variation in emission energies observed in nanorings. Optimising control of this parameter will allow for the tuning of nanoring optical properties.

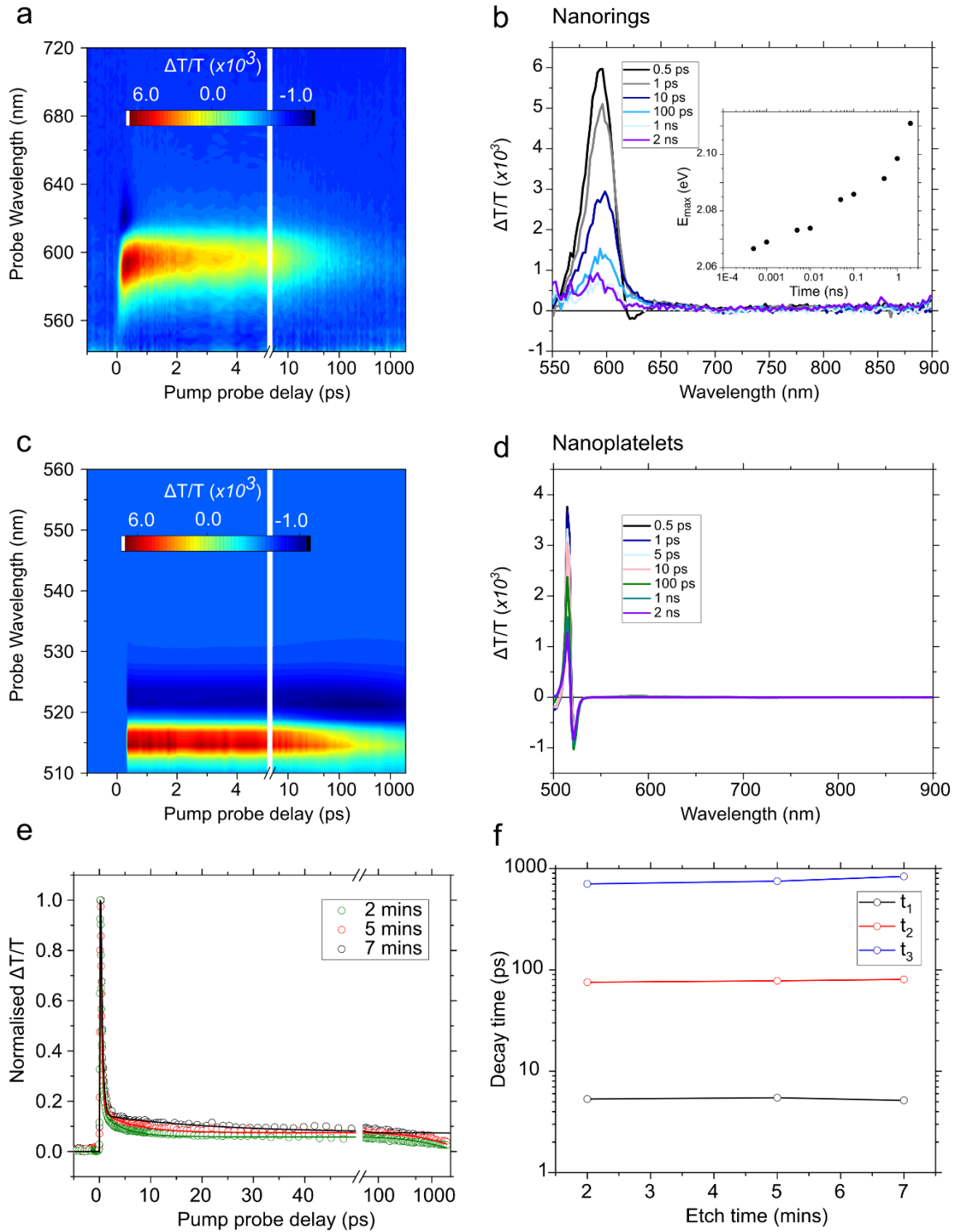

**Figure 4: Transient absorption spectrum of nanoplatelets and nanorings following photoexcitation with a sub-10 fs pulse. a.** Transient absorption spectrum of nanorings. There is a gradual red-shift (~50 meV; see inset) and narrowing of the bleach peak following photoexcitation which is ascribed to hopping of excitations between defect sites and to nanorings with the lowest bandgap within the ensemble. **b.** Selected spectral slices at given time delays of transient absorption

spectrum of nanorings. The weakly pronounced negative wing around ~625 nm, which is typically assigned to biexciton effects in nanocrystals,[51,52] suggests that exciton-exciton interactions are relatively weak in nanorings. **c.** Transient absorption spectrum of nanoplatelets. Following photoexcitation there is relatively little shift in the excitonic peak, with a significantly slower, multi-exponential decay as compared to nanorings. **d.** Selected spectral slices at given time delays of transient absorption spectrum of nanorings. The GSB is relatively long-lived as compared to in nanorings. **e.** Decay rate of the ground state bleach of nanorings as a function of etch time. **f.** The first and second decay constants ($t_1$ and $t_2$) do not change significantly with etch time. The final decay constant ($t_3$) appears to slightly lengthen as the platelet center is further etched.

Although several studies have addressed the steady-state properties of CdSe nanorings, no experiments have been able to elucidate how the exciton *dynamics* are influenced by their toroidal structure.[7,15] In order to directly probe the exciton dynamics in nanorings and understand how these excitons interact with phonon modes, we conducted femtosecond transient absorption (fs-TA) spectroscopy. A solution of nanorings was excited with a 9 fs pump pulse (upper limit verified by second-harmonic-generation frequency-resolved optical gating (SHG-FROG); **SI, S5**) centered at 530 nm such that it was partially resonant with the exciton transitions of the rings.[53] Figure 4a shows the transient absorption spectrum of nanorings, where $\Delta T/T$ is plotted as a function of probe wavelength and time delay between pump and probe. $\Delta T$ is the change in the transmission of the sample with/without the pump pulse, and $T$ is the transmission without the pump pulse.

The positive narrow signal centered at around 580 nm (Figure 4a,b) corresponds to the ground-state bleach (GSB) of nanorings, which agrees well with the steady-state absorption (Figure 1b; we note at the fluences used here there are ~0.15 excitations per nanorings; see **SI, S7**). This GSB decays with two lifetimes, an initial fast component of ~2 ps, followed by a much slower decay with a ~30 ps lifetime. The bleach red-shifts over a period of ~10 ps before beginning to decay (see Figure 4b inset). On the lower-energy edge of the bleach there is a weak negative feature. In nanoplatelets (Figure 4c,d) the transient absorption response is quite different. Here there is a relatively slow tri-exponential decay of the bleaching peak with an average decay time of ~920 ps and little-to-no-peak shifting following excitation (fit **SI, S7**). There is also a strong negative band on the red-edge of the GSB.

Several groups have performed and analyzed the transient absorption spectra of CdSe nanoplatelets. The derivative-like spectrum can be explained well by a model that accounts for state-filling-induced bleach of the electron-heavy hole and electron-light hole exciton transitions, a shift in the center frequency of these transitions due exciton-exciton interactions and bandgap renormalization.[52,54–57] Hole-trapping at surface defects and delocalization of excitons over the nanoplatelet have also been stated to occur in the first ~100 ps following photoexcitation in these materials.[58] This potentially explains why the GSB at 512 nm is comparatively short-lived compared to the negative feature at 525 nm, with state filling at the band edge replaced by trap state (*e.g.* at the surface[59]) absorption and a large Stark shift induced by more delocalized electrons at a band edge level (see **SI, S7**).

In nanorings there is a much weaker derivative-like shape to the spectrum with the negative wing on the red low-energy edge of the GSB decaying rapidly over ~1 ps. The spectrum is strongly reminiscent of that observed in metal halide perovskites, where following photoexcitation there is rapid (sub-500 fs) renormalization of the bandgap, giving rise to an early-time derivative feature, followed by cooling and hopping of excitations to localization sites.[60–62] The lack of persistence of this negative wing suggests that hot carrier induced exciton-exciton interactions are also weaker in nanorings in keeping with a picture of localized excitations.[63,64] This conclusion is supported by experiments on partially etched platelets, where the initial fast decay time remains independent of etch time, but the slow component lengthens, potentially as a result of more sub-bandgap trap states being introduced as the etch progresses (Figure 4e,f). This lack of negative feature in the transient absorption spectrum of nanorings compared to nanoplatelets also suggests a smaller biexciton binding energy in these materials. This is in line with the increased thickness of nanorings compared to nanoplatelets where Coulomb interactions are screened by the inorganic material.

Given the significant inhomogeneous broadening in the emission of nanorings, the red-shift (~50 meV) in the center frequency of the GSB over ~1 ps, may also represent energy transfer to the lowest energy nanorings within the solution ensemble with different recombination rates within the manifold.[65,66] However, the relatively large (~90 meV) Stokes shift between the absorption and emission of nanorings suggests that this may only be a partial effect. Indeed, performing pump-probe measurements as a function of ring concentration shows minimal change in the degree of red-shift (~35 meV in dilute solutions) suggesting that energy transfer does not play a dominant role (see **SI, S7** for further discussion and FRET calculations which show a FRET efficiency <2%). We note that in general, understanding the TA spectra of nanocrystals remains challenging, and the observations here will likely aid further interpretation of the TA spectra of CdX nanocrystals.

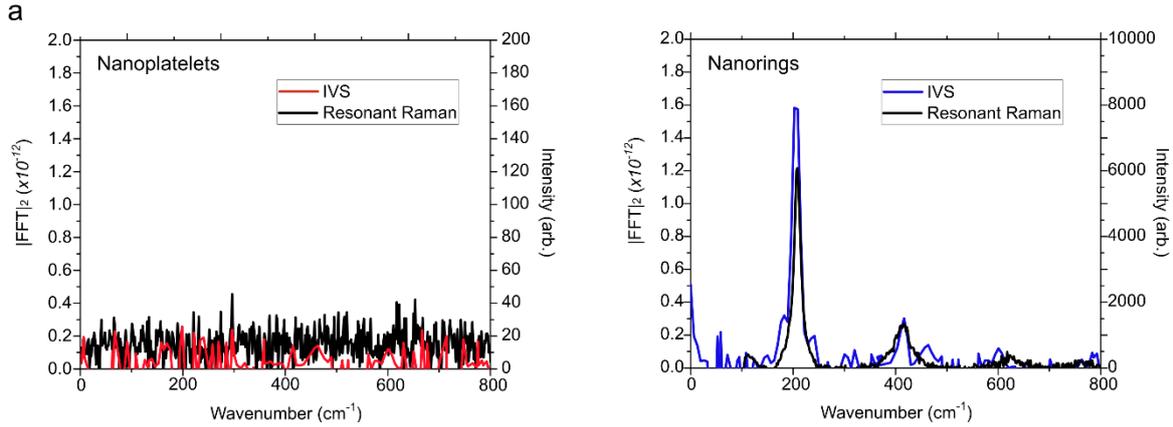

**Figure 5: Impulsive vibrational spectrum and resonant Raman spectra of nanorings and nanoplatelets. a.** Subtracting the electronic decay from the nanoring bleach and Fourier transforming the residual gives the impulsive (state-specific) vibrational spectrum (blue). Two clear peaks are observed at 200 cm$^{-1}$ and 400 cm$^{-1}$ corresponding to the CdSe LO-phonon stretch and its replica. These peaks also appear in the resonant (532 nm) Raman spectra (black). **b.** Repeating the same analysis for nanoplatelets appears to show no modes in either the IVS or Raman spectra, suggesting weaker exciton-phonon coupling in nanoplatelets compared to nanorings.

As the samples are excited with a sub-10 fs pump pulse, vibrational wavepackets can be generated on both the ground- and excited-state potential energy surfaces.[67–69] These then appear as an oscillatory feature on top of the electronic decay, as seen in the TA map and kinetics.[48] In order to extract the frequency of the vibrational modes, the electronic component of the TA spectrum in Figure 5a and b was subtracted and the remaining oscillatory component fast Fourier transformed into the frequency domain (see **SI, S8**; also includes wavelength resolved Fourier transform map). In the case of nanorings a single mode centered at ∼200 cm$^{-1}$ (frequency resolution 16 cm$^{-1}$; dephasing time ∼800 fs) is observed to modulate the GSB, whereas in nanoplatelets no modes are seen to modulate the spectra (Figure 5). Performing steady-state resonant (532 nm) and off-resonant (633 nm) Raman measurements also shows the 200 cm$^{-1}$ mode to be present in nanorings but there are no observable modes in the Raman spectra of nanoplatelets, consistent with previous studies where only weak-exciton phonon coupling has been observed.[70] Analyzing the literature shows this 200 cm$^{-1}$ mode corresponds to an LO-phonon of CdSe;[71–73] this mode also appears in the steady state and time domain Raman spectra of partially etched nanorings confirming it to be introduced by the etch, but is likely not necessarily a consequence of the annular shape (**SI, S9**). Strong coupling of excitons with this phonon mode further supports the idea of rapid localization of excitons in nanorings following photoexcitation. This is in contrast to the picture of excitons in nanoplatelets where the greater delocalization likely leads to weaker exciton-phonon coupling.[70,74–77] Based on our DFT calculations, the strong localization of the excitons in the nanorings can be attributed to some degree to the strong localization of the hole states that emerged in the nanoring electronic structure. The relatively low intensity of the phonon-replica at 400 and 800 cm$^{-1}$, suggest only a small difference in structure between the ground and excited states in nanorings (*i.e.* little lattice relaxation on photoexcitation). A full quantitative description involving Huang-Rhys parameters is beyond the scope of the present work but suggests the nanoring structure is not highly deformable by the soft LO-phonon mode.[78]

We note that in the case of nanoplatelets on increasing both the pump fluence and energy there is a qualitative shortening of the GSB lifetime (SI, S6). This is in keeping with a model of state filling within a valence band where the cooling time decreases when pumping closer to the band edge and Auger recombination increases with greater excitation densities.[79,80] In contrast, nanorings do not appear to show any pump energy or excitation density dependence in their transient absorption response. This would again fit with a trap-like picture where excitons hop from site-to-site before localizing, likely assisted by the 200 cm$^{-1}$ phonon mode, as opposed to being delocalized in band states (see SI, S6 for further discussion).

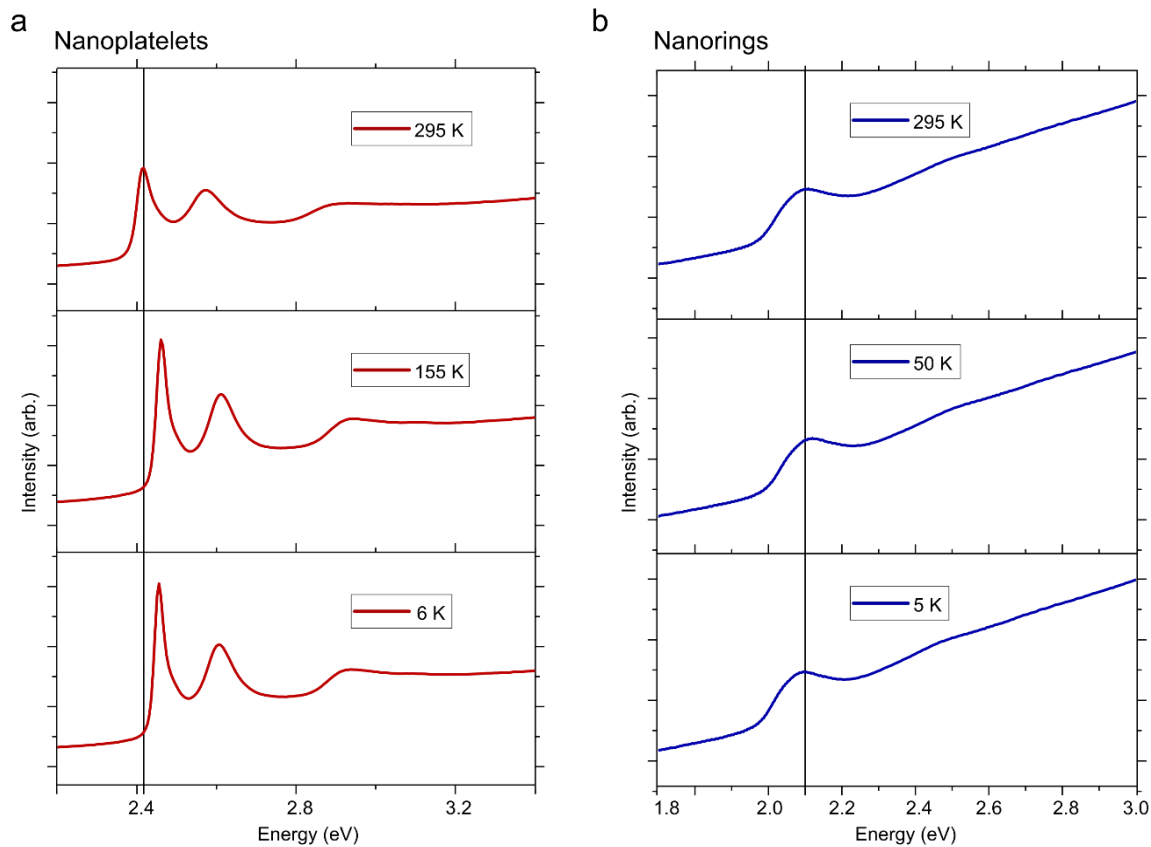

**Figure 6: Temperature dependent absorption spectra of nanoplatelets and nanorings. a.** The absorption spectra of nanoplatelets (red) initially blue-shifts to higher energies on cooling, followed by a gradual red-shift when cooling below 20 K. Solid line marks absorption maximum at 295 K. **b.** In nanorings (blue) there is a similar blue- and then red-shift in the absorption maximum on cooling.

Temperature X-ray diffraction measurements confirm there is no phase transition on cooling for either nanorings or nanoplatelets (see **SI, S3**).

Figure 6 reports temperature dependent absorption measurements for films of nanoplatelets and nanorings. In the case of both nanorings and nanoplatelets the absorption spectrum initially shows a concomitant blue-shift to higher energies on cooling. In both materials the absorption spectrum then begins to red-shift to lower energies below a certain temperature (~60 K in nanoplatelets and ~100 K in nanorings). In nanoplatelets this effect has been previously ascribed to phonon-mediated thermal redistribution between two levels having different oscillator strengths (**SI, S10**).[81] Given the similar behaviour we can hence tentatively suggest in this regard a similar electronic state picture potentially exists for excitons in nanorings. Here, when the thermal energy becomes smaller than any energy splitting, only the lower level will be thermally populated. However, at elevated temperatures both upper and lower states are populated and because the upper state has a higher oscillator strength, this state will dominate the optical behaviour (see further discussion below). In nanorings (and to a weaker extent in nanoplatelets) there is a broad sloping sub-band gap signal between 1.8-2 eV (nanorings). Studies have shown that unpassivated selenium on the surface of nanocrystals can result in the formation of midgap charge carrier trapping states.[82,83] Given the poorly-passivated, selenium-rich surface of nanorings it is likely these along with additional states, closely spaced at the localized valence band edge (Figure 2) contribute to this absorption. Inhomogeneous broadening, from variations in ring thickness and diameter, also may lower the conduction band in nanorings resulting in transitions to sub-band gap states;[84] in nanoplatelets the uniform quantum confinement suggests this to be a less of an effect. From PDS measurements (Figure 1d) we can estimate the absorption coefficient of the sub-band gap states. For nanorings an $\alpha$ ~100 cm$^{-1}$ is found (1.5-1.8 eV); in nanoplatelets $\alpha$ ~400 cm$^{-1}$ (1.7-2 eV). As a result of this low oscillator strength we do not expect direct absorption from these states to be significant.

In nanoplatelets a similar behaviour as to that discussed for absorption measurements is also realised in the emission spectra, with a non-monotonic change in the PL intensity and wavelength with temperature, and a red-shifting zero-phonon line below 80 K (Figure 7a). A second weakly emissive state also appears centered around 530-550 nm in nanoplatelets. We note this state is unlikely to arise from a population of thicker nanoplatelets, as the thickest stable CdSe nanoplatelets synthesised thus far (6 monolayers) have their emission centered around ~590 nm and signatures of such nanoplatelets would also be expected to be present in the absorption spectra.[85] For the case of nanorings the temperature resolved emission shows some similarities to nanoplatelets but also some differences. For example, where the red-shift in the nanoplatelets PL is ~10 meV between 50K and 4K, in nanorings the red-shift is ~25-30 meV in the temperature range 150 K to 4 K. In the former, this shift is ascribed to thermalization of excitons into the lowest lying dark exciton (F) sub-level, with ~10 meV representing the bright-dark exciton energy splitting,[81] although mixing between sub-levels also influences this gap.[83] The similarity in PL response between nanorings and nanoplatelets at first suggests a three-level bright, dark and ground-state exciton picture, with an albeit larger energy gap between the bright and dark exciton states. However, in this picture, the PL decay is expected to be faster as the temperature is lowered, because there is rapid relaxation into the dark exciton state (and diminished radiative decay from bright excitons). In nanoplatelets this is indeed observed,[47,86] but in nanorings the response is more complex (see **SI, S11**). The 0-1.5 ns PL dynamics in a 4-150 K range can be well described with a bi-exponential decay. Initially between 150-80 K the fast component of this decay quickens on cooling, while the slower component lengthens. Below 80 K the trend is then reversed, *i.e.* fast component increases again, (80 K, $t_{fast}$ = 0.03 ns, $t_{slow}$ = 0.57 ns; 4 K, $t_{fast}$ = 0.09 ns, $t_{slow}$ = 0.49 ns). The relative weight of the fast to slow component drops from ~2 at 160 K to ~0.5 at 50 K before increasing to ~2 again at 4 K (see **SI, S11**). This complex transient PL response suggests a model consisting solely of a bright-dark exciton doublet is not suitable to describe the electronic structure of nanorings.[27] This is consistent with DFT calculations (Figure 2) which suggest the valence band maximum to be a localized hole state. This state potentially mixes with bright and dark exciton states, shifting their relative energetic separation and population. Accessing a full picture of this is beyond the scope of this work and would likely require additional magneto-optical measurements, but also shows how the electronic structure of these nanoparticles sits between 0D and 1D confined particles. Hence, in general, our results suggest that between 160-80 K there is thermal population of bright, dark and any trap states (see below trap activation energy ~8 meV ($\equiv$ 92 K)), the increase in decay time on cooling within this range suggests radiative recombination from a bright state dominates. Below 80 K thermalization into a dark manifold occurs. However, the overall non-monotonic changes in lifetime suggest the mixing with the localized valence band states creates a complex electronic response. Indeed, if nanorings followed the same electronic structure as platelets the magnitude of the red-shift might be expected to be even smaller, due to the suggested increased thickness of nanorings.[86]

In nanorings the PL intensity is also rapidly reduced on decreasing the temperature with a second spectrally broad (~70 meV at 4 K) emissive state (centered around 740 nm) growing in intensity on cooling. The difference in energy between trap states in nanoplatelets and nanorings suggests that although they are both likely located in a similar location *i.e.* at the surface, the change in electronic structure arising from shape nanocrystal shape alters their energies. At 5 K the main excitonic emission and emission from this second state are almost equal in intensity, in contrast to the nanoplatelets where emission from the excitonic state is ~10 times greater than emission from the state at ~530-550 nm, even at 4 K. We assign these observations to population competition with low-lying trap states as has been observed in 0D quantum dots.[87,88] At low temperatures excitations localize in these trap states from which some fraction of the excitons can decay radiatively. The PL intensity of the band-edge (BE) emissive state and trap state will be proportional to the occupation, *n*, which follows a Boltzmann distribution, giving:

$$\frac{PL_{BE}}{PL_{trap}} \propto \frac{n_{BE}}{n_{trap}} = \frac{g_{BE}}{g_{trap}} e^{\left(\frac{-\Delta E}{kT}\right)} \text{ (Equation 1)}$$

In this equation ΔE is the energy difference between traps and the band-edge excitonic state and $g_{BE}$ and $g_{trap}$ are the band edge exciton and trap state degeneracy respectively. Fitting the PL ratio intensities from these two states as shown in Figure 5 for nanoplatelets and nanorings then allows us to determine the trap-state depth. In this fitting we make the assumption that any charge transfer is sufficiently fast to establish an equilibrium between traps and band-edge states and consequently only perform the fit at high temperatures (T>200 K).[89] Fitting the data yields trap activation energies of 21 ± 8 meV for nanoplatelets and 8 ± 6 meV in nanorings. We note these are both consistent with the observation of room temperature emission from trap states. One final observation from these low temperature PL measurements is that the PL maximum initially slightly blue-shifts and then begins to red-shift below 150 K in nanorings, suggesting the main emission from the main excitonic state is still consistent with a model similar to that observed in nanoplatelets (Figure 7c and e). As a final point we note that the exact position of the emission maximum in nanorings varies between batches (5-10 meV); the trends in PL peaks and quantitative values extracted *e.g.* trap activation energies remain unaltered with the error bars on these values taking this into account.

In 0D CdSe (and CdS or CdTe) quantum dots, temperature-dependent emission measurements have revealed trap activation energies of a similar magnitude to those reported here for platelets and rings at 7-50 meV.[90–92] Adding a shell to these materials can greatly increase the activation barrier for traps, *e.g.* 180 meV in CdSe/CdS/ZnS QDs,[93] although in some cases the shell has been shown to also introduce new shallow trap states.[94] The trap activation energies reported in 0D PbS quantum dots have tended to be larger at ~30-50 meV and as high as 180 meV in QD solids.[89,95] However, in general the trap activation barrier in 0D QDs is dependent on a number of factors beyond shape, *e.g.* surface passivation, size, *etc*.[96,97] This suggests that the increased surface area to volume ratio of platelets and rings, as compared to 0D dots, does not necessarily make them more susceptible to shallow traps and that similar surface treatments are required to eliminate trapping.

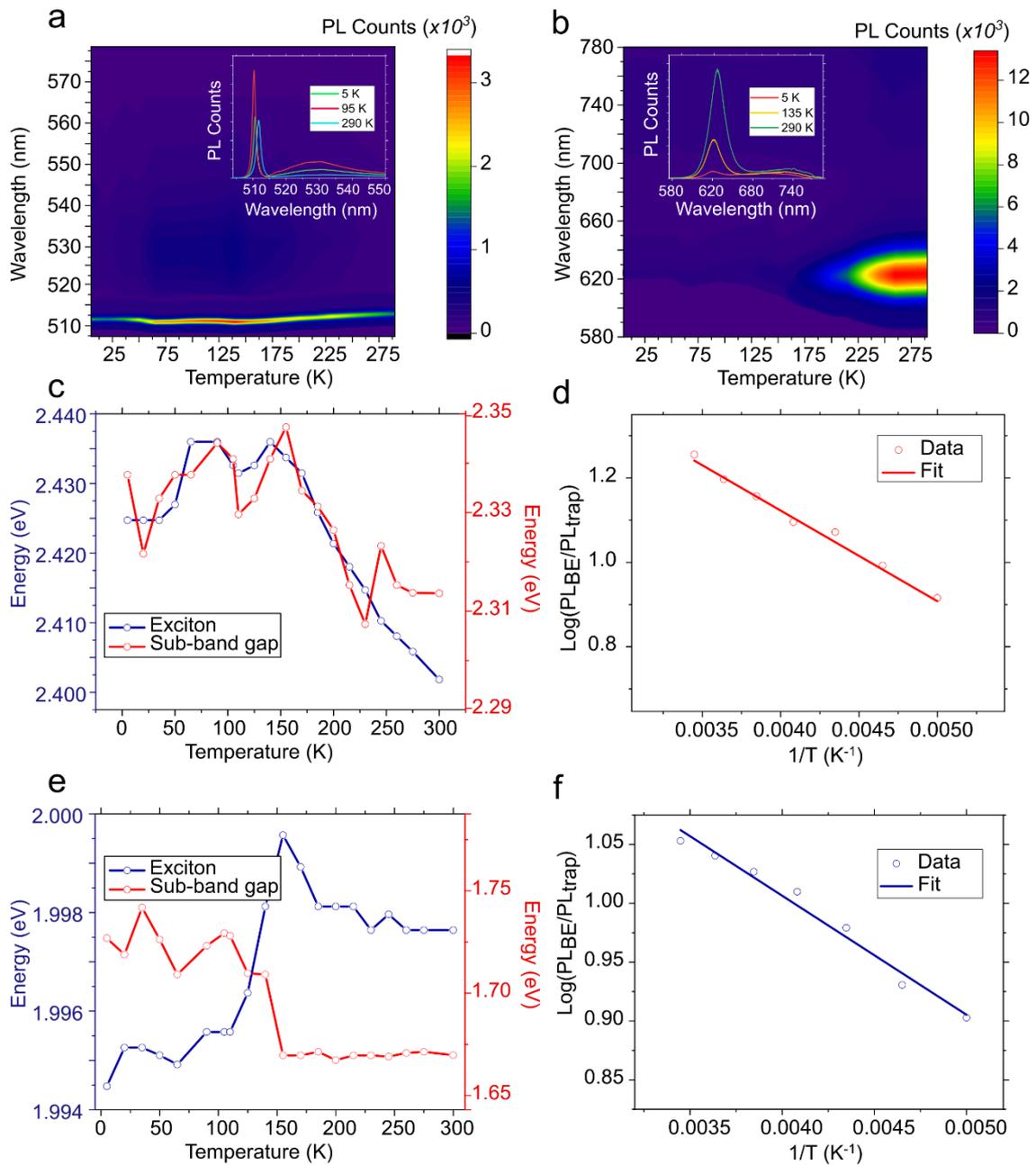

**Figure 7: Low temperature emission spectra of CdSe nanoplatelets and nanorings. a.** Emission spectra of CdSe nanoplatelets excited at 405 nm. A non-monotonic change in the PL intensity and maximum peak position is observed on cooling, with some secondary sub-band gap emission centered around 530-550 nm. Inset shows selected spectra. **b.** Emission spectra of CdSe nanorings. As for nanoplatelets, the excitonic emission initially blue-shifts to higher energies on cooling before red-shifting when below 100 K. There is a secondary sub-band gap emission peak which grows in intensity on cooling. There is a secondary sub-band gap emission peak which grows in intensity on cooling. Inset

shows selected spectra. The data between 700-780 nm is taken using a second grating; insets show spectral slices. **c.** Maximum peak position for excitonic and sub-band gap emission as a function of temperature for CdSe nanoplatelets. The maximum peak position for both the sub-band gap and excitonic state in nanoplatelets initially shows a red-shift followed by a blue-shift below 100 K. Blue axis corresponds to exciton emission; red axis is sub-band gap emission. **d.** Log of ratio of integrated intensities for excitonic and sub-band gap emission in nanoplatelets. At room temperature there is ~16 times more emission from the excitonic state as compared to sub-band gap emission. At low temperature this ratio is reduced to ~8 (**SI, S9**). Fitting the high temperature data ($T > 200$ K) with the model in equation 1 allows extraction of the trap activation energy of ~21 meV. **e.** Maximum peak position for excitonic and sub-band gap emission as a function of temperature for CdSe nanorings. Where the sub-band gap emission only blue-shifts to higher energies on cooling, the excitonic emission in nanorings initially shows a red-shift followed by a blue-shift. Blue axis corresponds to exciton emission; red axis is sub-band gap emission. **f.** Log of ratio of integrated intensities for excitonic and sub-band gap emission in nanorings ($PL_{BE}/PL_{trap}$). At room temperature there is ~11 times more emission from the excitonic state as compared to sub-band gap emission, whereas at low temperature the emission intensity from these two states is almost equal (**SI, S9**). Fitting the high temperature data ($T > 200$ K) with the model in equation 1 allows extraction of the trap activation energy of ~8 meV.

Broad red-shifted emission with respect to a main exciton peak has, in 0D nanocrystals, been attributed to surface emission. This surface emission is particularly prominent in small nanocrystals with a large surface-area-to-volume ratio, where emission from the surface states is coupled electronically to the excitonic states.[98–101] Given the large surface-area-to-volume ratio of nanoplatelets and nanorings we investigated whether semi-classical electron transfer models,[98–102] typically used to ascribe core and surface emission, could be applied here. At high temperatures (200-300 K) we find this 'coupled surface and core' model fits well (**SI, S11**). However, at low temperatures, the model deviates significantly from the experimental PL behaviour. From the overall fit we can we extract a free energy difference ($\Delta G$) between the core and surface states of $25 \pm 2$ meV in both nanoplatelets and nanorings, similar to values found in 0D CdSe systems.[98–101] The general deviation of the model from the data might be because of the difference in electronic structure between nanoplatelets/nanorings (*e.g.* localized valence band states in nanorings) and the 0D dots this model is normally applied to. In 0D dots the model assumes a single exciton state is coupled to a single surface state *via* a low frequency mode representing the bath and a high frequency LO-phonon mode of the nanocrystal.[99] The LO-phonon mode allows tunneling between core and surface states, but in nanorings/nanoplatelets several modes or intermediate states might be involved which the model is unable to capture. We do indeed observe a relative increase in the surface emission at low temperatures and the energy difference between the exciton and surface emission is significantly larger than $k_B$T at these temperatures, suggesting population exchange cannot be purely thermal.[100] We note previous studies on nanoplatelets have suggested that surface emission arises from a broad distribution of states within the bandgap (deep traps). In nanoplatelets the emission red shifted from the main excitonic peak is then ascribed to recombination between trapped holes and delocalized electrons.[30,103,104] In nanorings further work is required to understand if such a picture also holds. Single-nanoring PL measurements shown in Figure 3 are unable to resolve PL from the trap state, likely due to its extremely weak emission. Understanding whether this broad emission also exists at the single-particle level, as well as improved models to capture any coupling between exciton and surface potential energy surfaces, should be a focus of future work.

In order to understand the loss of emission towards trap states in nanorings and nanoplatelets at *room temperature* transient photoluminescence measurements were performed (Figure 8) on solutions of nanorings and nanoplatelets. In both nanoplatelets and nanorings a multi-exponential decay is observed. In nanoplatelets the initial fast component of the decay (>10 ns) has been ascribed to cooling, energy transfer and Auger decay which are accelerated due to a 'giant oscillator strength' effect.[2] The delayed component of the emission (which has been shown to extend to beyond 20 μs) has been assigned to reversible charge trapping/detrapping. The emission decay in platelets is uniform across the band in keeping with the low-inhomogeneous broadening and a picture of delocalized excitations. By contrast the PL decay in nanorings is much shorter than platelets, although also multi-exponential, with the decay here well described with three decay constants of 1.3 ± 0.2 ns, 7.2 ± 2 ns and 40.3 ± 2 ns (**see SI, S12** for all decays and associated weighting coefficients). The absence of a long-lived emission component in the decay suggests a model using reversible-carrier trapping/delayed luminescence cannot be used to explain the PL decay in these materials.[105] In light of the aforementioned evidence for localization of excitations in nanorings, the fast component of the PL decay is potentially associated with a combination of cooling processes and energy transfer between different nanorings within the inhomogeneously broadened PL. The slower part of the decay may then reflect the time taken for the excitations to hop and localize between sites on a particular nanoring. Once the carriers are localized the recombination is relatively fast likely as a result of the strong coupling with CdSe LO-phonons and negatively charged selenium rich surface sites (**SI, S4**), hence explaining the relatively overall short lifetime (and low PL quantum yield).[48,106,107]

The nanoring PL decay is not uniform across the band, with faster decay on the blue high-energy edge, and with the PL peak maximum shifting to lower energies over ~1 ns. This is again likely a consequence of inhomogeneous broadening within the emission of nanorings, with some energy transfer to nanorings with a smaller band gap following photoexcitation. There could also be energy transfer between different sites of varying thickness (and hence band gap) on the individual nanorings. The low overall FRET efficiency between nanorings (see **SI, S7**) will further serve to reduce the PL quantum yield. However, the stability of the single nanoring PL precludes transient photoluminescence measurements to fully characterize this and establish whether the low quantum yield for nanorings results from lower yield for all nanorings, or arises because some are emissive and others are quenched. The results are also consistent with the hypothesis that rings which have a higher band energy (bluer PL) undergo faster radiative recombination. From our DFT simulations (Figure 2) it would be expected that these nanorings have a larger radius. Finally, we note in both platelets and nanorings emission from the sub-gap states at 530-550 nm and 740 nm (see Figure 8 inset for spectral slices) are relatively long-lived compared to emission from excitons.

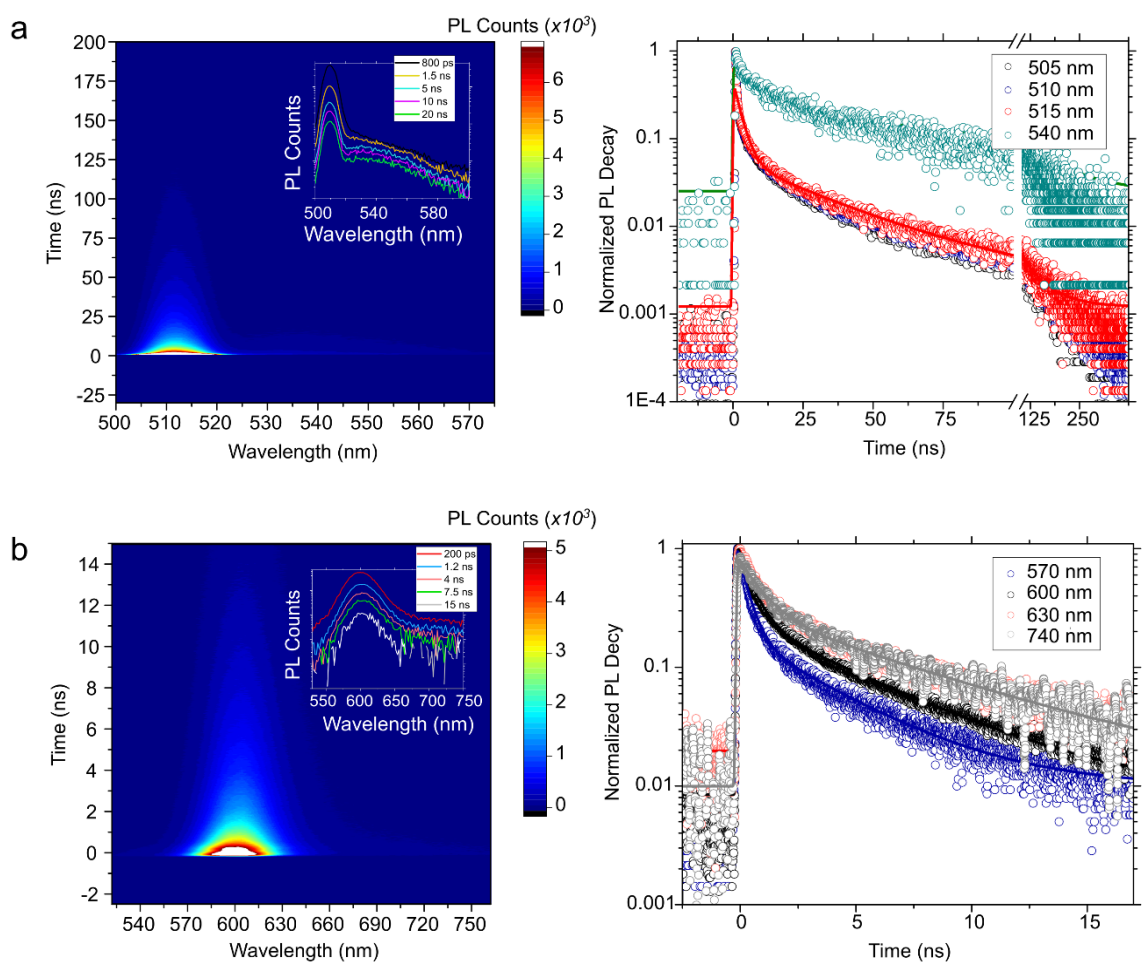

**Figure 8: Transient photoluminescence measurements of nanoplatelets and nanorings excited at 405 nm: a.** Transient photoluminescence spectrum of 4-monolayer nanoplatelets in solution. There is a multi-exponential decay (right) which decays uniformly across the emission band. At early times the decay is attributed to energy transfer and cooling whereas at later times corresponds to reversible charge trapping. A second, long lived emission feature at ~540 nm is observed. Inset shows spectral slices at selected time delays. **b.** Transient photoluminescence spectrum of nanorings. The emission is significantly shorter lived than nanoplatelets with a faster decay at higher energies (left). An additional emissive feature ~740 nm is observed which has relatively long lived emission compared with that of excitons and is suggested to originate from sub-band gap defect states.

The PL decay of nanoplatelets and nanorings can also be fit to more complex stretched exponential functions (see **SI, S13**) which allow estimation of the number of traps, trapping and detrapping rates and energy gap between trap sites.[108,109] Fitting the decay in Figure 8 to such models, shows that the average number of traps in nanorings is 3 times higher than in nanoplatelets with trap activation energies of similar magnitude to those found with the fitting of Equation 1 (see **SI, S13**). In nanorings the trapping rate is almost 100 times faster than the detrapping rate, whereas in nanoplatelets the two rates are comparable. This is again in keeping with observations from steady-state PL. Additionally, the stretching factors from these fits are consistently greater than 0.5, suggesting that FRET-based PL quenching by *e.g.* stacked nanoplatelets/nanorings is not significant.[110] Stacking between nanoplatelets might be expected to increase the efficiency of FRET due to the small Stokes shift between absorption and PL. Stacking has also been shown to be responsible for additional PL peaks in the low temperature emission spectra of nanoplatelets due to excimer formation;[111] although this is still somewhat debated.[86,112] In this work, we observe little impact of stacking on the exciton dynamics of nanoplatelets, *e.g.* no closely spaced PL peaks in low temperature emission spectra, potentially due to the relatively irregular shape of nanoplatelets synthesised (Figure 1a). From qualitative assessment of TEM images (**SI, S1**) nanorings do not appear to stack to as great an extent as nanoplatelets. The relatively large (~90 meV) Stokes shift between absorption and emission in these materials means the efficiency of nanoring-to-nanoring FRET is low. Performing concentration-dependent transient PL measurements shows no change in the PL lifetime on increasing the concentration (**SI, S13**), further suggesting any stacking does not play an important role in the observations here, *e.g. via* the formation of an indirect exciton between stacked nanorings which would be expected to have an increased lifetime. The shape and peak positions in the steady-state PL spectrum similarly do not vary with concentration (**SI, S13**). Furthermore, The energy difference between the excitonic PL and PL attributed to defects is also too large (~0.3 eV) to be as a consequence of excimer formation.[111] As a result, we can conclude stacking has little effect on the observations reported. In summary a variety of models can be used to fit the nanoplatelet and nanorings PL. These models typically only apply in the case of a two-level bright-dark exciton for II-VI colloidal system. In the case of nanorings this picture does not necessarily hold, hence the quantitative values should be treated with caution. The reasonable fit and derived parameters from these models however demonstrates that many of the models developed to understand trap states in QDs can be applied to nanocrystals of more unusual geometries.

Finally, we note atmospheric (gas exposure/light soaking) Kelvin probe measurements additionally reveal poor trap passivation of nanorings as compared to nanoplatelets (SI, Figure S14).[113,114]. This results in stronger and more irreversible oxygen binding to the surface of nanorings as compared to nanoplatelets, and suggests at least for nanoplatelets there are excess electrons at the surface and oxygen reduces the sensitivity of holes to surface defects. These results highlight the relation between the shape and surface (trap) passivation, suggesting in these more complex geometries new strategies or ligands might be required to passivate possibly inaccessible facets. (SI, Figure S14).[113,114] A larger effect was observed from atmospheric exposure in nanorings compared to nanoplatelets, and this was attributed to less effective passivation by surface ligands, allowing oxygen to have a greater effect. This highlights the importance of surface trap engineering when colloidal nanocrystals become modified into nontrivial topologies and we expect core-shell structures to become integral to these systems. Excess electrons at the nanoplatelet surface and reduces the sensitivity of holes to surface defects. We emphasise that for other measurements in this manuscript (transient absorption, photoluminescence, *etc.*) the samples were encapsulated (films) or sealed (solution) in an inert nitrogen or argon environment.

Conclusions

We have investigated exciton dynamics in colloidal CdSe nanorings. Using ultrasensitive, scatter-free absorption measurements, we have shown forming these rings from nanoplatelets introduces relatively little disorder into the materials. DFT simulations show the changes in the electronic structure lead to a red-shift in the absorption on etching the central hole of the platelets to form a ring. Variation in the size, thickness, and localization properties of the rings leads to a broad absorption and emission with a large Stokes shift as confirmed by single-particle PL measurements. Following photoexcitation, excitons on the nanoring hop between (surface) trap sites before localising, this is in contrast to the delocalized, band-like picture of excitations in nanoplatelets. These increased exciton localization likely emerges from the localized hole state due to the change in topology and electronic structure from the creation of the nanoring. There is also likely energetic hopping between nanorings of different band energies within the ensemble. The surfaces of nanorings have a large negative charge, suggesting the traps consist of regions of unpassivated selenium. In contrast to nanoplatelets, there is strong coupling between excitons and a LO-phonon mode at 200 cm$^{-1}$, which we suggest plays a key role in localising excitations. However, the shape does not necessarily appear to play a significant role in altering the exciton-phonon coupling as one might expect. *In-situ* characterization of nanorings during their formation from nanoplatelets suggests this enhanced exciton-phonon coupling is introduced during the etching process. Using low-temperature absorption and emission, we confirm this equilibrium between trap and excitonic emission, with the former dominating at low temperatures. Transient measurements as a function of temperature further demonstrate that despite a similar steady-state PL response to nanoplatelets the band-edge electronic structure in nanorings cannot be treated simply in terms of a bright-dark doublet. Analysis of steady state and transient PL data gives a semi-quantitative picture of surface traps and free energies in both nanoplatelets and nanorings and suggests potential couplings between the exciton and surface states.

These results have several implications for the engineering of 2D colloidal CdSe nanorings. Firstly, they suggest that passivation of negative selenium anion traps is key. This might be achieved either *via* appropriate surface ligands, core-shell structures, or by better removing excess selenium from the etch process. Given exciton-phonon scattering is a major PL loss pathway in these materials, our results additionally suggest that tuning of the material structure is required. This could perhaps be achieved by alloying the material,[115] to reduce coupling with phonon modes and alleviate strain in nanorings, an approach which has had success in some 2D transition metal dichalcogenides.[116,117] In general, the nature of trap states in these materials remains poorly understood and future work should aim to better characterize these, *e.g.*, *via* computational modelling. Here, we have shown that the application of models initially developed for 0D QDs can be applied to 2D and ring topologies for quantitative analysis of surface properties; these potentially challenge the idea of deep traps in these systems, but further work is required still to understand this.[100,104] From a synthetic viewpoint, ensuring all nanorings have the same diameter would lead to emission with narrower linewidths. This could potentially be achieved by flow-synthesis routes,[118] although we highlight in our methods key requirements in terms of original nanoplatelet size, aspect ratio and surface ligands for efficient and clean single ring formation through the etch method. We expand on the synthetic methods of Fedin *et al.* which may be applied to other material systems. One avenue of future study is the synthesis of colloidal topological nanostructures of semiconductors such as HgTe which is a topological insulator.[119,120] The expansive techniques applied here may provide a basis for understanding the exciton dynamics within topological systems of these different materials.

As active materials for quantum dot LEDs, the strong exciton-phonon coupling in nanorings is not advantageous, due to the asymmetry and PL broadening introduced.[121] However, these materials could find use in polariton lasers, where the strong exciton-phonon coupling increases the rate of polariton relaxation allowing for low lasing thresholds.[122–124] The fact that the emission is sharp (FWHM μeV range) while at the few-particle level means they could be promising candidates for photonic applications which can exploit their toroidal geometry.

For application in solar cells, the high trap density and the fact that the main carriers are excitons as opposed to free charges, limits the use of nanorings, although these materials would be optimal over nanoplatelets. However, if the PL quantum yield can be enhanced significantly using approaches proven for colloidal quantum dots, nanorings might find effective use in luminescent solar concentrators, where a large Stokes shift between absorption and emission coupled with the potential for highly directed emission is desirable.

In terms of applications that exploit the nanoring shape, our results suggest that surface traps introduced by etching of the platelet center might be a limiting factor.[16] However, our work points to a potentially unusual excitonic fine structure where there is an interplay between effects observed in 2D platelets and 0D dots. Investigating this further with magneto-optical spectroscopies should be the focus of future work.

Methods

CdSe Nanorings synthesis

CdSe nanorings (NRs) were synthesized according to the method of Fedin *et al.* following the chemical etching of CdSe nanoplatelets (NPLs) synthesized according to the method of Ithurria *et al.* All materials were purchased from Sigma-Aldrich unless otherwise stated and were used as received. A typical procedure is as follows.[7,2]

Cadmium myristate was prepared by first reacting myristic acid (1.37 g, 6 mmol) with sodium hydroxide (0.24 g, 6 mmol) in 240 mL methanol. Cadmium nitrate (0.617 g, 2 mmol) dissolved in 40 mL methanol was then added dropwise with vigorous stirring over 1 hour. Dropwise addition along with vigorous stirring are important for complete reaction and purity of cadmium myristate. The white precipitate was collected through vacuum filtration and washed 3 times with 30 mL methanol. The powder was dried under vacuum overnight at 80 °C.

A 50 mL 3-necked flask was loaded with cadmium myristate (170 mg) and 1-octadecene (ODE, 15 mL) then degassed at 100 °C for 10 minutes, forming a colourless solution. The flask was switched to $N_2$ then cooled to room temperature. Cadmium myristate is insoluble in ODE below 100 °C and will precipitate to form a cloudy suspension if left at room temperature. Selenium powder (12 mg, 100-mesh) was added under flowing $N_2$ then the flask degassed at 90 °C for 30 minutes. The flask was switched to $N_2$ and heated to 140 °C until the Se had fully dissolved. Larger Se loadings increased the product yield but also increased the likelihood of secondary nucleation of different thickness platelets. Waiting for full dissolution of the Se below CdSe nucleation temperature assisted the reproducibility of the synthesis by minimising differences in reaction kinetics during the temperature ramp phase. Proceeding with the temperature ramp before full dissolution of the Se powder may yield a monodisperse platelet thickness, however this becomes more difficult especially at higher concentrations of Se and cadmium acetate. Increased sensitivity to Se powder particle size (mesh) and Cd acetate addition temperature were observed with incomplete dissolution of Se, resulting in polydisperse thicknesses and nanocube side products. A separate receiving Schlenk flask containing methylcyclohexane (MCH, 10 mL) and degassed oleic acid (OA, 2 mL) in $N_2$ was prepared. The 3-necked flask was rapidly heated to 240 °C, and finely ground cadmium acetate dihydrate (40 mg, 0.15 mmol) was added to the flask under nitrogen flow when the temperature reached 190 °C. Fast addition of powder was easily achieved by transferring the pre-ground powder to a tubular metal or anti-static plastic spatula. Environmental humidity was an unexpected factor in this step. Bertrand *et al.* demonstrated control of nanoplatelet aspect ratio to water present in the reaction.[31] It was discovered higher aspect ratio nanoplatelets were produced with increased humidity in winter, leading to nanorings with multiple perforations. To modify the aspect ratio, cadmium acetate could be added as a mix of the anhydrous and dihydrate forms, with a 50:50 molar ratio being successful for low aspect ratio platelets in wetter conditions. The solution was maintained 240 °C for 4 minutes. The exact reaction time could be between 1 minute and 8 minutes, with longer times associated to an extent with larger lateral dimensions. Short reaction times resulted in reduced nanoplatelet yields. The solution turned turbid upon successful growth of nanoplatelets whilst remaining an orange-red colour. However, if the colour changed to darker red, this indicated nucleation of nanocubes or dots as a side product. The hot solution was then extracted with a glass syringe and metal needle and rapidly injected into the receiving flask. For the extraction, a 30 mL glass syringe with 14 or 16 gauge stainless steel needle was used to expedite the transfer time. The needle and syringe was purged with nitrogen using the receiving flask, also pre-perforating the septum to facilitate rapid insertion. The cloudy solution was allowed to settle for 3 hours then transferred to an Ar glovebox then centrifuged at 12000 g for 5 minutes. The supernatant was discarded and the precipitate was resuspended in 4 mL MCH and filtered through a 0.2 μm PTFE syringe filter, to remove large aggregates. The crude solution was purified twice by precipitation with ethanol, centrifugation and resuspension in MCH. The purified NPLs were stored in MCH. NPLs were colloidally stable in MCH but would eventually precipitate in hexane.

A quantity of purified nanoplatelets was dispersed into degassed ODE (3 mL) and degassed OAm (1 mL). These were stored as three separate stock solutions in the glovebox and could be mixed together before transferring to the flask. The absorption spectrum of the NPLs in MCH was used to determine the amount of stock solution to be used. When the sharp absorption feature at 511 nm for a solution of NPLs diluted by a factor of 300 was 0.2 A in a 1 cm cuvette, 1 mL of the undiluted solution was used for the ring etch procedure. A convenient method of dilution was to disperse 6.6 μL of stock solution into 2 mL additional solvent. The quantity of NPL stock was varied linearly according to absorbance, thus for a diluted stock measured at 0.4 A, one should use 0.5 mL of NPL stock solution. The NPL solution in ODE/OAm was transferred to a 25 mL 3-necked flask purged with $N_2$. The solution was heated to 80 °C under $N_2$ for 15 minutes to evaporate the MCH then the flask was slowly degassed whilst being allowed to cool to room temperature. The heating allowed faster and more complete removal of excess MCH solvent. Selenium powder (Alfa-Aesar, 7.9 mg, 325-mesh) was dispersed and sonicated in degassed OAm (1 mL) and 0.2 mL of this suspension was added to the degassed NPL solution. The finer powder mesh was used to achieve more reproducible results, allowing for better initial dispersion in OAm, and to improve dissolution rate when reaching solubility temperature. Selenium may also be pre-dissolved in OAm by heating the suspension on a hotplate within the glovebox at 180 °C overnight, resulting in an orange-brown solution to be used instead of the powder suspension. Vortexing the powder suspension with fast extraction before sedimentation was useful in ensuring a correct amount of Se was used. The flask was degassed at room temperature for 10 minutes before heating to 140 °C under $N_2$. The temperature was maintained at 140 °C for 10 minutes, then tri-n-butylphosphine (0.2 ml) was injected into the flask and the temperature was increased to 220 °C. A visible darkening of the solution was observed during the temperature plateau, which lightened following injection of tri-n-butylphosphine, darkening again upon heating to 220 °C. The heating mantle was removed immediately upon reaching 220 °C and the flask was allowed to cool to room temperature. At 37 °C, finely ground cadmium formate (MP Biomedicals, 10 mg) was added to the flask, and the mixture was stirred for 1 hour. The solution was transferred to an Ar glovebox and purified by precipitation with a mixture of ethanol/acetone, centrifugation and resuspension in hexane.

Bright-field Transmission Electron Microscopy

Bright-field transmission electron microscopy was performed with a FEI Tecnai F20 TEM at 200 kV operating voltage. Samples were prepared by deposition of diluted nanocrystals in hexane onto carbon-coated Cu grids (Agar AGS160).

Photoluminescence Quantum Efficiency (PLQE)

Photoluminescence quantum efficiency was performed on a home-built measurement setup consisting of an integrating sphere (Labsphere 4P-GPS-053-SL), collection fibre (Andor SR-OPT8019), spectrograph (Andor Kymera-328i) and detector (Andor iDus 420). The setup was calibrated for spectral sensitivity with a NIST-traceable quartz-tungsten-halogen lamp (Newport 63967-200QC-OA). Excitation was performed with a 520 nm temperature-controlled diode laser (Thorlabs). PLQE values were determined *via* the integrating sphere method.[125]

Absorption Spectroscopy

Linear absorption spectra of colloidal nanoplatelets solutions, placed in a 1 mm path length cuvette (Hellma), were measured using a commercial PerkinElmer Lambda 750 UV–vis–NIR setup equipped with a 10 cm integrating sphere module attachment. A Xe lamp was used as an excitation source, and all measurements were performed under standard ambient conditions. In order to collect the scattered light, the sample cuvette was placed on the front window of the sphere. The spectra were measured simultaneously with the solvent hexane to correct for its absorption.

Single particle photoluminescence spectroscopy

The samples were prepared by dropcasting a diluted solution of nanorings (dilution up to 5000 times) onto ~ 5 x 5 mm glass slides. They were then mounted on the cold finger of a cryostat (Oxford Instrument) allowing the control of the temperature from ≈ 4 K to room temperature. The samples were excited with a continuous wave diode-pumped solid state laser (Thorlabs DJ532) emitting at 533 nm, mounted in its temperature controlled laser mount (Thorlabs TCLDM9). The excitation was focused using a microscope objective (NA = 0.6, spot size ≈ 1 μm). The incident power density was kept around 5 μW/μm². The luminescence was collected using the same optic and spectrally analyzed with an ACTON SP2760i Roper Scientific-Princeton Instruments spectrometer coupled to a nitrogen-cooled SPEC10-2KB-LN (RS-PI) CCD camera (1200 lines per mm grating).

Temperature-Dependent Absorption

An Agilent Cary 6000i UV–Vis–NIR spectrophotometer with blank substrate correction was used. Spin-coated samples on fused silica substrates were placed in a continuous-flow cryostat (Oxford Instruments Optistat CF-V) under a helium atmosphere. We allowed the sample temperature to equilibrate for 30 min before taking data.

DFT Calculations

All DFT calculations were performed using the Quantum Espresso suite (v6.4).[126,127] We used the SG15 norm-conserving pseudopotentials generated by ONCVPSP and the electronic wavefunctions were expanded in a plane wave basis with an energy cutoff of 60 Ry.[128] The exchange correlation functional was approximated by the Perdew-Burke-Ernzerhof (PBE) generalized gradient approximation[129]. A vaccum spacing of 15 Å was added to the supercell in the *z*-dimension to remove any spurious interactions, and atomic positions were relaxed until the residual forces were <0.01 eV/Å. The frequency-dependent dielectric function was computed using the independent particle approximation.

Bulk zincblende CdSe was used to create nanoplatelets with their top and bottom surfaces being the {100} facets, which were passivated by Cl ligands. The nanorings were created in a 3 ×3 supercell, and all structures were stoichiometric in Cd, Se and Cl composition. The Monkhorst-Pack *k*-point sampling scheme used for Brillouin zone has divisions of less than 0.03 Å$^{-1}$, and only the Γ point was sampled in the *z*-direction. [130]

Photothermal Deflection Spectroscopy

PDS measures the refractive index change due to heat that is caused by nonradiative relaxation when the incoming light is absorbed and can be considered a scatter-free technique for measuring absorption capable of measuring 5–6 orders of magnitude weaker absorption than the band-edge absorption. For the measurements, a monochromatic pump light beam produced by a combination of a Light Support MKII 100 W xenon arc source and a CVI DK240 monochromator was shone on the sample (film on Spectrosil fused silica substrate), inclined perpendicular to the plane of the sample, which on absorption produces a thermal gradient near the sample surface *via* nonradiative relaxation induced heating. This results in a refractive index gradient in the area surrounding the sample surface. This refractive index gradient was further enhanced by immersing the sample in an inert liquid FC-72 Fluorinert (3M Company) that has a high refractive index change *per* unit change in temperature. A fixed-wavelength CW laser probe beam, produced using a Qioptiq 670 nm fiber-coupled diode laser with temperature stabilizer for reduced beam pointing noise, was passed through this refractive index gradient, producing a deflection proportional to the absorbed light at that particular wavelength, which was detected by a differentially amplified quadrant photodiode and lock-in amplifier (Stanford Research SR830) combination. Scanning through different wavelengths gives the complete absorption spectra.

Kelvin Probe

Kelvin Probe measurements were measured using a RHC Kelvin Probe System (RHC KP030, KP Technology Ltd.), which measured the contact potential difference between the sample surface and a cylindrical gold tip (4 mm diameter). The samples were prepared under dry $N_2$ conditions and transferred to the environmental chamber containing the Kelvin Probe. The work function (WF) measurement started immediately after the installation, when the samples were conditioned under dry $N_2$. After stabilization of the work-function signal, indicating reaching the steady-state we began tracking of the WF value. A surface spectroscopy module (AM0.5, SPS020, KP Technology), was used for surface photovoltage (SPV) measurements.

Temperature dependent X-Ray diffraction

XRD was performed using a Bruker X-ray D8 Advance diffractometer with Cu Kα radiation (λ = 1.541 Å) with Johansson monochromators to eliminate Kα2. Low-temperature measurements were made on cooling between 300 and 12 K using an Oxford Cryosystem PheniX stage (50 K steps; 20 min waited for equilibration of temperature). Spectra were collected with an angular range of 10° < 2θ <

60° and Δθ = 0.014 31° over 60 min. Measurements were made on drop-casted films from the nanoplatelet or nanoring suspension onto precleaned silicon substrates.

Picosecond Transient Absorption

The picosecond transient absorption (ps-TA) experiments were performed using an Yb-based amplified system (PHAROS, Light Conversion) providing 14.5 W at 1030 nm and 38 kHz repetition rate. The probe beam is generated by focusing a portion of the fundamental in a 4 mm YAG substrate and spans from 520 to 900 nm. The pump beam is generated by seeding a portion of the fundamental to a narrow band optical parametric oscillator (ORPHEUS-LYRA, Light Conversion). The pump pulse was set to 500 nm. The sample solutions were placed in 1 mm path length cuvettes (Hellma). The pump and probe beams were focused to a size of 280 μm × 240 μm and 55 μm × 67 μm, respectively. The pump fluence was typically 30 μJ/cm$^2$. The white light is delayed using a computer-controlled piezoelectric translation stage (Newport), and a sequence of probe pulses with and without the pump is generated using a chopper wheel (Thorlabs) on the pump beam. The probe pulse transmitted through the sample was detected by a silicon photodiode array camera (Stresing Entwicklungsbüro; visible monochromator 550 nm blazed grating).

Transient Photoluminescence

To record the time-resolved emission scan or photoluminescence decay of the samples, time-correlated single-photon counting (TCSPC) was performed. Samples were excited with a pulsed laser (PicoQuant LDH400 40 MHz) 470 nm, with the resulting photoluminescence decay collected on a 500 mm focal length spectrograph (Princeton Instruments, SpectraPro2500i) with a cooled CCD camera. The instrument response was determined by scattering excitation light into the detector using a piece of frosted glass; a value of 265 ps was obtained.

Low temperature transient PL spectroscopy

Films of nanorings drop-cast on glass slides were mounted on the cold finger of a cryostat (Oxford Instrument) allowing the control of the temperature from ≈ 4 K to 293 K. A C5680 Hamamatsu streak camera, incorporating a M5675 synchro-scan unit, coupled to an Acton SP2760i spectrometer was used. The streak camera was synchronized with the pulse train of the excitation. In the experiment, pulses of $tr_{es}$ ≈ 2 ps are delivered by a picosecond Ti:Sapphire laser operating at 82 MHz. Frequency conversion is achieved in a BBO crystal to reach an excitation wavelength of $\lambda_{ex}$ = 460 nm. The time resolution is less than 20 ps. The excitation was focused using a microscope objective (NA = 0.6, spot size ≈ 1 μm) and the excitation power density was kept around 1 μW/μm². The luminescence was then collected using the same objective.

Femtosecond Transient Absorption Spectroscopy

The fs-TA experiments were performed using an Yb-based amplified system (Pharos, Light Conversion) providing 14.5 W at 1030 nm and a 38 kHz repetition rate. The probe beam was generated by focusing a portion of the fundamental in a 4 mm YAG substrate and spanned from 520 to 1400 nm. The pump beam was generated in a home-built noncollinear optical parametric (NOPAs; 37° cut BBO, type I, 5° external angle) pumped with either the second or third harmonic of the source. The NOPAs output (~4–5 mW power) was centered at either 520, 660, or 860 nm, and pulses were compressed using a chirped mirror wedge prism (Layterc) combination to a temporal duration of 12 and 17 fs, respectively (upper limit determined by SHG-FROG). The white light was delayed using a computer-controlled piezoelectric translation stage, and a sequence of probe pulses with and without pump was generated using a chopper wheel on the pump beam. The pump irradiance was at 19 μJ/cm². After the sample, the probe pulse was split with a 950 nm dichroic mirror (Thor Laboratories). The visible light (520–950 nm) was then imaged with a silicon photodiode array camera (Stresing Entwicklunsbüro; visible monochromator 550 nm blazed grating) with the near-infrared proportion of the probe seeded to an IR monochromator (1200 nm blazed grating) and imaged using an InGaAs photodiode array camera (Sensors Unlimited). This technique allows simultaneous collection of the entire probe spectrum in a single shot. Offsets for the differing spectral response of the detectors was accounted for in the post-processing of data.

Steady-state Raman

Raman measurements were performed with a Renishaw inVia Raman microscope under ambient conditions. Excitation was provided by 532 nm and 633 nm laser lines. The Raman emission was collected by a Lecia 100× objective (N.A. = 0.85) and dispersed by a 1800 lines per mm grating. Measurements were performed on thin films of nanoplatelets and nanorings atop a glass microscope slide.

Zeta-potential measurements

Zeta-potential and size distribution measurements were performed on a Malvern Zetasizer NanoZSP, equipped with a 633 nm He-Ne laser with a maximum power of 10 mW.  A glass cuvette (PCS1115) was loaded with 400 μL of QDs in hexane (1 mg/mL), and a Malvern Universal Dip Cell (ZEN1002)was used for measurement of zeta-potential**.** All measurements were performed at 25 °C, with a minimum equilibration period of 2 minutes following sample loading. Zeta-potential distributions were averaged over 5 measurements, each of which consisted of between 50 to 100 runs.

## Associated Content

## Supporting Information

TEM images of partially etched nanoplatelets, DFT calculations, temperature-dependent X-Ray diffraction measurements, optical pulse compression, analysis of impulsive vibrational spectroscopy and pump-probe data, Raman spectra of partially etched nanorings, temperature dependent absorption and emission spectra, zeta potential measurements.

The Supporting Information is available free of charge at [url to be added in proof].

The raw data underlying this manuscript is freely available at https://doi.org/10.17863/CAM.58042.

## Acknowledgements

We thank the EPSRC (programme grant: EP/M005143/1) and Winton Program for Physics of Sustainability for financial support. R.H.F. and Y.L. acknowledge support from the Simons Foundation (grant 601946). This work was performed using resources provided by the Cambridge Service for Data Driven Discovery (CSD3) operated by the University of Cambridge Research Computing Service (www.csd3.cam.ac.uk). R.P. thanks Ryan Brady (Cambridge) for assistance with zeta-potential and DLS measurements, Gianluca Grimaldi (Cambridge) for assistance with interpretation of transient absorption data, Juan Climente (Universitat Jaume I, Spain) for insightful discussions and H.P., S.P. and T.P. for support. J.X thanks Simon Dowland (Cambridge) for assistance with PLQE measurements and X. Hong (Reading) for useful discussions. E. R. and S. D. S. acknowledge the European Research Council (ERC) under the European Union's Horizon 2020 research and innovation program (HYPERION, grant agreement number 756962). E. R. was partially supported by an EPSRC

Departmental Graduate Studentship. S. D. S. acknowledges funding from the Royal Society and Tata Group (UF150033).

References

(1) Smith, A. M.; Nie, S. Semiconductor Nanocrystals: Structure, Properties and Band Gap Engineering. *Acc. Chem. Res.* **2010**, *43*, 190–200.

(2) Ithurria, S.; Tessier, M. D.; Mahler, B.; Lobo, R. P. S. M.; Dubertret, B.; Efros, A. L. Colloidal Nanoplatelets with Two-Dimensional Electronic Structure. *Nat. Mater.* **2011**, *10*, 936–941.

(3) Milliron, D.; Hughes, S. M.; Cui, Y.; Manna, L.; Li, J.; Wang, L. W.; Alivisatos, A. P. Colloidal Nanocrystal Heterostructures with Linear and Branched Topology. *Nature* **2004**, *430*, 190–195.

(4) Sigle, D. O.; Zhang, L.; Ithurria, S.; Dubertret, B.; Baumberg, J. J. Ultrathin CdSe in Plasmonic Nanogaps for Enhanced Photocatalytic Water Splitting. *J. Phys. Chem. Lett.* **2015**, *6*, 1099–1103.

(5) Li, Q.; Zhou, B.; McBride, J. R.; Lian, T. Efficient Diffusive Transport of Hot and Cold Excitons in Colloidal Type-II CdSe/CdTe Core/Crown Nanoplatelet Heterostructures. *ACS Energy Lett.* **2017**, *2*, 174–181.

(6) Rowland, C. E.; Susumu, K.; Stewart, M. H.; Oh, E.; Mäkinen, A. J.; O'Shaughnessy, T. J.; Kushto, G.; Wolak, M. A.; Erickson, J. S.; L. Efros, A.; A.; Huston, A. L.; Delehanty, J. B. Electric Field Modulation of Semiconductor Quantum Dot Photoluminescence: Insights into the Design of Robust Voltage-Sensitive Cellular Imaging Probes. *Nano Lett.* **2015**, *15*, 6848–6854.

(7) Fedin, I.; Talapin, D. V. Colloidal CdSe Quantum Rings. *J. Am. Chem. Soc.* **2016**, *138*, 9771–9774.

(8) Mitchell, B.; Massey, W. S. Algebraic Topology: An Introduction. *Am. Math. Mon.* **1968**, *75*, 427-428.

(9) Li, E.; Eggleton, B. J.; Fang, K.; Fan, S. Photonic Aharonov-Bohm Effect in Photon-Phonon Interactions. *Nat. Commun.* **2014**, *5*, 3225.

(10) Bayer, M.; Korkusinski, M.; Hawrylak, P.; Gutbrod, T.; Michel, M.; Forchel, A. Optical Detection of the Aharonov-Bohm Effect on a Charged Particle in a Nanoscale Quantum Ring. *Phys. Rev. Lett.* **2003**, *90*, 186801.

(11) Chen, P.; Whaley, K. B. Magneto-Optical Response of CdSe Nanostructures. *Phys. Rev. B - Condens. Matter Mater. Phys.* **2004**, *70*, 045311.

(12) Ju, S.; Jeong, S.; Kim, Y.; Watekar, P. R.; Han, W. T. Demonstration of All-Optical Fiber Isolator Based on a CdSe Quantum Dots Doped Optical Fiber Operating at 660 nm. *J. Light. Technol.* **2013**, *31*, 3093–3098.

(13) Baghramyan, H. M.; Barseghyan, M. G.; Kirakosyan, A. A.; Ojeda, J. H.; Bragard, J.; Laroze, D. Modeling of Anisotropic Properties of Double Quantum Rings by the Terahertz Laser Field. *Sci. Rep.* **2018**, *8*, 6145.

(14) Tadić, M.; Aukarić, N.; Arsoski, V.; Peeters, F. M. Excitonic Aharonov-Bohm Effect: Unstrained *versus* Strained Type-I Semiconductor Nanorings. *Phys. Rev. B - Condens. Matter Mater. Phys.* **2011**, *84*, 125307.

(15) Hartmann, N. F.; Otten, M.; Fedin, I.; Talapin, D.; Cygorek, M.; Hawrylak, P.; Korkusinski, M.; Gray, S.; Hartschuh, A.; Ma, X. Uniaxial Transition Dipole Moments in Semiconductor Quantum Rings Caused by Broken Rotational Symmetry. *Nat. Commun.* **2019**, *10*, 3253.

(16) Meinardi, F.; Colombo, A.; Velizhanin, K. A.; Simonutti, R.; Lorenzon, M.; Beverina, L.; Viswanatha, R.; Klimov, V. I.; Brovelli, S. Large-Area Luminescent Solar Concentrators Based on Stokes-Shift-Engineered Nanocrystals in a Mass-Polymerized PMMA Matrix. *Nat. Photonics* **2014**, *8*, 392–399.

(17) Gerislioglu, B.; Ahmadivand, A.; Pala, N. Hybridized Plasmons in Graphene Nanorings for Extreme Nonlinear Optics. *Opt. Mater. (Amsterdam, Neth.).* **2017**, *73*, 729-735.

(18) Flatten, L. C.; Christodoulou, S.; Patel, R. K.; Buccheri, A.; Coles, D. M.; Reid, B. P. L.; Taylor, R. A.; Moreels, I.; Smith, J. M. Strong Exciton-Photon Coupling with Colloidal Nanoplatelets in an Open Microcavity. *Nano Lett.* **2016**, *16*, 7137–7141.

(19) Wu, J.; Shao, D.; Li, Z.; Manasreh, M. O.; Kunets, V. P.; Wang, Z. M.; Salamo, G. J. Intermediate-Band Material Based on GaAs Quantum Rings for Solar Cells. *Appl. Phys. Lett.* **2009**, *95*, 071908.

(20) Guo, Q.; Kim, S. J.; Kar, M.; Shafarman, W. N.; Birkmire, R. W.; Stach, E. A.; Agrawal, R.; Hillhouse, H. W. Development of CuInSe2 Nanocrystal and Nanoring Inks for Low-Cost Solar Cells. *Nano Lett.* **2008**, *8*, 2982-2987.

(21) Kanté, B.; Park, Y. S.; O'Brien, K.; Shuldman, D.; Lanzillotti-Kimura, N. D.; Jing Wong, Z.; Yin, X.; Zhang, X. Symmetry Breaking and Optical Negative Index of Closed Nanorings. *Nat. Commun.* **2012**, *3*, 1180.

(22) Shalaev, V. M. Optical Negative-Index Metamaterials. *Nat. Photonics*. **2007**, *1*, 41-48.

(23) Caglar, M.; Pandya, R.; Xiao, J.; Foster, S. K.; Divitini, G.; Chen, R. Y. S.; Greenham, N. C.; Franze, K.; Rao, A.; Keyser, U. F. All-Optical Detection of Neuronal Membrane Depolarization in Live Cells Using Colloidal Quantum Dots. *Nano Lett.* **2019**, *19*, 8539-8549.

(24) Xiong, S.; Xi, B.; Qian, Y. CdS Hierarchical Nanostructures with Tunable Morphologies: Preparation and Photocatalytic Properties. *J. Phys. Chem. C* **2010**, *114*, 14029-14035.

(25) Cohen Stuart, M. A. Supramolecular Perspectives in Colloid Science. *Colloid Polym. Sci..* **2008**, *286*, 855-864.

(26) Földi, P.; Kálmán, O.; Benedict, M. G.; Peeters, F. M. Networks of Quantum Nanorings: Programmable Spintronic Devices. *Nano Lett.* **2008**, *8*, 2556-2558.

(27) Nirmal, M.; Norris, D. J.; Kuno, M.; Bawendi, M. G.; Efros, A. L.; Rosen, M. Observation of the Dark Exciton in CdSe Quantum Dots. *Phys. Rev. Lett.* **1995**, *75*, 3728.

(28) Kuroda, T.; Mano, T.; Ochiai, T.; Sanguinetti, S.; Sakoda, K.; Kido, G.; Koguchi, N. Optical Transitions in Quantum Ring Complexes. *Phys. Rev. B - Condens. Matter Mater. Phys.* **2005**, *72*, 205301.

(29) Warburton, R. J.; Schäflein, C.; Haft, D.; Bickel, F.; Lorke, A.; Karrai, K.; Garcia, J. M.; Schoenfeld, W.; Petroff, P. M. Optical Emission from a Charge-Tunable Quantum Ring. *Nature* **2000**, *405*, 926–929.

(30) Ithurria, S.; Bousquet, G.; Dubertret, B. Continuous Transition From 3D to 1D Confinement Observed during the Formation of CdSe Nanoplatelets. *J. Am. Chem. Soc.* **2011**, *133*, 3070–3077.

(31) Bertrand, G. H. V.; Polovitsyn, A.; Christodoulou, S.; Khan, A. H.; Moreels, I. Shape Control of Zincblende CdSe Nanoplatelets. *Chem. Commun.* **2016**, *52*, 11975–11978.

(32) Biadala, L.; Shornikova, E. V.; Rodina, A. V.; Yakovlev, D. R.; Siebers, B.; Aubert, T.; Nasilowski, M.; Hens, Z.; Dubertret, B.; Efros, A. L.; Bayer, M. Magnetic Polaron on Dangling-Bond Spins in CdSe Colloidal Nanocrystals. *Nat. Nanotechnol.* **2017**, *12*, 569–57.

(33) Mooney, J.; Kambhampati, P. Get the Basics Right: Jacobian Conversion of Wavelength and Energy Scales for Quantitative Analysis of Emission Spectra. *J. Phys. Chem. Lett.* **2013**, *4*, 3316–3318.

(34) Diroll, B. T.; Chen, M.; Coropceanu, I.; Williams, K. R.; Talapin, D. V.; Guyot-Sionnest, P.; Schaller, R. D. Polarized Near-Infrared Intersubband Absorptions in CdSe Colloidal Quantum Wells. *Nat. Commun.* **2019**, *10*, 4511.

(35) Ekimov, A. I.; Hache, F.; Ricard, D.; Flytzanis, C.; Polytechnique, E.; Minchen, T. U. Absorption and Intensity-Dependent Photoluminescence Measurements on CdSe Quantum Dots: Assignment of the First Electronic Transitions. *J. Opt. Soc. Am. B* **1993**, *10*, 100–107.

(36) Kim, D.; Kim, D. H.; Lee, J. H.; Grossman, J. C. Impact of Stoichiometry on the Electronic Structure of PbS Quantum Dots. *Phys. Rev. Lett.* **2013**, *110*, 1–5.

(37) Giansante, C.; Infante, I. Surface Traps in Colloidal Quantum Dots: A Combined Experimental and Theoretical Perspective. *J. Phys. Chem. Lett.* **2017**, *8*, 5209–5215.

(38) Shornikova, E. V.; Golovatenko, A. A.; Yakovlev, D. R.; Rodina, A. V.; Biadala, L.; Qiang, G.; Kuntzmann, A.; Nasilowski, M.; Dubertret, B.; Polovitsyn, A.; Moreels, I. Surface Spin Magnetism Controls the Polarized Exciton Emission from CdSe Nanoplatelets. *Nat. Nanotechnol.* **2020**, *15,* 277-282.

(39) Momma, K.; Izumi, F. VESTA: A Three-Dimensional Visualization System for Electronic and Structural Analysis. *J. Appl. Crystallogr.* **2008**, *41*, 653–658.

(40) Jackson, W. B.; Amer, N. M.; Boccara, A. C.; Fournier, D. Photothermal Deflection Spectroscopy and Detection. *Appl. Opt.* **1981**, *20*, 1333.

(41) Tessier, M. D.; Javaux, C.; Maksimovic, I.; Loriette, V.; Dubertret, B. Spectroscopy of Single CdSe Nanoplatelets. *ACS Nano* **2012**, *6*, 6751–6758.

(42) Turtos, R. M.; Gundacker, S.; Omelkov, S.; Mahler, B.; Khan, A. H.; Saaring, J.; Meng, Z.;

Vasil'ev, A.; Dujardin, C.; Kirm, M.; Moreels, I. On the Use of CdSe Scintillating Nanoplatelets as Time Taggers for High-Energy Gamma Detection. *npj 2D Mater. Appl.* **2019**, *3*, 37. https://doi.org/10.1038/s41699-019-0120-8.

(43) Greeff, C. W.; Glyde, H. R. Anomalous Urbach Tail in GaAs. *Phys. Rev. B* **1995**, *51*, 1778-1783.

(44) Ikhmayies, S. J.; Ahmad-Bitar, R. N. A Study of the Optical Bandgap Energy and Urbach Tail of Spray-Deposited CdS:In Thin Films. *J. Mater. Res. Technol.* **2013**, *2*, 221–227.

(45) Guyot-Sionnest, P.; Lhuillier, E.; Liu, H. A Mirage Study of CdSe Colloidal Quantum Dot Films, Urbach Tail, and Surface States. *J. Chem. Phys.* **2012**, *137*, 154704.

(46) Tessier, M. D.; Spinicelli, P.; Dupont, D.; Patriarche, G.; Ithurria, S.; Dubertret, B. Efficient Exciton Concentrators Built From Colloidal Core/Crown CdSe/CdS Semiconductor Nanoplatelets. *Nano Lett.* **2014**, *14*, 207–213.

(47) Biadala, L.; Liu, F.; Tessier, M. D.; Yakovlev, D. R.; Dubertret, B.; Bayer, M. Recombination Dynamics of Band Edge Excitons in Quasi-Two-Dimensional CdSe Nanoplatelets. *Nano Lett.* **2014**, *14*, 1134–1139.

(48) Pandya, R.; Chen, R. Y. S.; Cheminal, A.; Dufour, M.; Richter, J. M.; Thomas, T. H.; Ahmed, S.; Sadhanala, A.; Booker, E. P.; Divitini, G.; Deschler, F.D,; Greenham, N. C.; Ithurria, S.; Rao, A. Exciton-Phonon Interactions Govern Charge-Transfer-State Dynamics in CdSe/CdTe Two-Dimensional Colloidal Heterostructures. *J. Am. Chem. Soc.* **2018**, *140*, 1409–14111.

(49) Bezinge, L.; Maceiczyk, R. M.; Lignos, I.; Kovalenko, M. V.; DeMello, A. J. Pick a Color MARIA: Adaptive Sampling Enables the Rapid Identification of Complex Perovskite Nanocrystal Compositions with Defined Emission Characteristics. *ACS Appl. Mater. Interfaces* **2018**, *10*, 18869-18878.

(50) Li, B.; Brosseau, P. J.; Strandell, D. P.; Mack, T. G.; Kambhampati, P. Photophysical Action Spectra of Emission from Semiconductor Nanocrystals Reveal Violations to the Vavilov Rule Behavior from Hot Carrier Effects. *J. Phys. Chem. C* **2019**, *123*, 5092-5098.

(51) Klimov, V. I.; Multiexciton Phenomena in Semiconductor Nanocrystals, In *Nanocrystal Quantum Dots*, 2; Klimov, V. I.; CRC Press: Boca Raton, Fla **2017**, Chapter 5, pp 147-213.

(52) Kambhampati, P. Multiexcitons in Semiconductor Nanocrystals: A Platform for Optoelectronics at High Carrier Concentration. *J. Phys. Chem. Lett.* **2012**, *3*, 1182–1190.

(53) Trebino, R.; DeLong, K. W.; Fittinghoff, D. N.; Sweetser, J. N.; Krumbügel, M. A.; Richman, B. A.; Kane, D. J. Measuring Ultrashort Laser Pulses in the Time-Frequency Domain Using

Frequency-Resolved Optical Gating. *Rev. Sci. Instrum.* **1997**, *68*, 3277–3295.

(54) Wu, K.; Li, Q.; Jia, Y.; McBride, J. R.; Xie, Z. X.; Lian, T. Efficient and Ultrafast Formation of Long-Lived Charge-Transfer Exciton State in Atomically Thin Cadmium Selenide/Cadmium Telluride Type-II Heteronanosheets. *ACS Nano* **2015**, *9*, 961–968.

(55) Wu, K.; Li, Q.; Du, Y.; Chen, Z.; Lian, T. Ultrafast Exciton Quenching by Energy and Electron Transfer in Colloidal CdSe Nanosheet-Pt Heterostructures. *Chem. Sci.* **2015**, *6*, 1049–1054.

(56) Kunneman, L. T.; Schins, J. M.; Pedetti, S.; Heuclin, H.; Grozema, F. C.; Houtepen, A. J.; Dubertret, B.; Siebbeles, L. D. A. Nature and Decay Pathways of Photoexcited States in CdSe and CdSe/CdS Nanoplatelets. *Nano Lett.* **2014**, *14*, 7039-7045.

(57) Li, Q.; Xu, Z.; McBride, J. R.; Lian, T. Low Threshold Multiexciton Optical Gain in Colloidal CdSe/CdTe Core/Crown Type-II Nanoplatelet Heterostructures. *ACS Nano* **2017**, *11*, 2545–2553.

(58) Morgan, D. P.; Maddux, C. J. A.; Kelley, D. F. Transient Absorption Spectroscopy of CdSe Nanoplatelets. *J. Phys. Chem. C* **2018**, *122*, 23772–23779.

(59) Kambhampati, P. On the Kinetics and Thermodynamics of Excitons at the Surface of Semiconductor Nanocrystals: Are There Surface Excitons? *J. Chem. Phys.* **2015**, *446*, 92-107.

(60) Yang, Y.; Ostrowski, D. P.; France, R. M.; Zhu, K.; Van De Lagemaat, J.; Luther, J. M.; Beard, M. C. Observation of a Hot-Phonon Bottleneck in Lead-Iodide Perovskites. *Nat. Photonics* **2016**, *10*, 53–59.

(61) Price, M. B.; Butkus, J.; Jellicoe, T. C.; Sadhanala, A.; Briane, A.; Halpert, J. E.; Broch, K.; Hodgkiss, J. M.; Friend, R. H.; Deschler, F. Hot-Carrier Cooling and Photoinduced Refractive Index Changes in Organic-Inorganic Lead Halide Perovskites. *Nat. Commun.* **2015**, *6*¸ 8420.

(62) Manser, J. S.; Kamat, P. V. Band Filling with Free Charge Carriers in Organometal Halide Perovskites. *Nat. Photonics* **2014**, *8*, 737–743.

(63) Geiregat, P.; Houtepen, A.; Justo, Y.; Grozema, F. C.; Van Thourhout, D.; Hens, Z. Coulomb Shifts upon Exciton Addition to Photoexcited PbS Colloidal Quantum Dots. *J. Phys. Chem. C* **2014**, *118*, 22284–22290.

(64) Kambhampati, P. Multiexcitons in Semiconductor Nanocrystals: A Platform for Optoelectronics at High Carrier Concentration. *J. Phys. Chem. Lett.* **2012**, *3*, 1182–1190.

(65) Moroz, P.; Royo Romero, L.; Zamkov, M. Colloidal Semiconductor Nanocrystals in Energy Transfer Reactions. *Chem. Commun.* **2019**, *55*, 3033–3048.

(66) De Weerd, C.; Gomez, L.; Zhang, H.; Buma, W. J.; Nedelcu, G.; Kovalenko, M. V.; Gregorkiewicz, T. Energy Transfer between Inorganic Perovskite Nanocrystals. *J. Phys. Chem. C* **2016**, *120*, 13310–13315.

(67) Liebel, M.; Kukura, P. Broad-Band Impulsive Vibrational Spectroscopy of Excited Electronic States in the Time Domain. *J. Phys. Chem. Lett.* **2013**, *4*, 1358–1364.

(68) Mooney, J.; Saari, J. I.; Myers Kelley, A.; Krause, M. M.; Walsh, B. R.; Kambhampati, P. Control of Phonons in Semiconductor Nanocrystals *via* Femtosecond Pulse Chirp-Influenced Wavepacket Dynamics and Polarization. *J. Phys. Chem. B* **2013**, *117*, 15651–15658.

(69) Sagar, D. M.; Cooney, R. R.; Sewall, S. L.; Kambhampati, P. State-Resolved Exciton-Phonon Couplings in CdSe Semiconductor Quantum Dots. *J. Phys. Chem. C* **2008**, *112*, 9124–9127.

(70) Maddux, C. J. A.; Kelley, D. F.; Kelley, A. M. Weak Exciton-Phonon Coupling in CdSe Nanoplatelets from Quantitative Resonance Raman Intensity Analysis. *J. Phys. Chem. C* **2018**, *122*, 27100–27106.

(71) Lin, C.; Kelley, D. F.; Rico, M.; Kelley, A. M. The Surface Optical Phonon in CdSe Nanocrystals. *ACS Nano* **2014**, *8*, 3928–3938.

(72) Mork, A. J.; Lee, E. M. Y.; Tisdale, W. A. Temperature Dependence of Acoustic Vibrations of CdSe and CdSe-CdS Core-Shell Nanocrystals Measured by Low-Frequency Raman Spectroscopy. *Phys. Chem. Chem. Phys.* **2016**, *18*, 28797–28801.

(73) Dzhagan, V. M.; Valakh, M. Y.; Milekhin, A. G.; Yeryukov, N. A.; Zahn, D. R. T.; Cassette, E.; Pons, T.; Dubertret, B. Raman-and IR-Active Phonons in CdSe/CdS Core/Shell Nanocrystals in the Presence of Interface Alloying and Strain. *J. Phys. Chem. C* **2013**, *117*, 18225–18233.

(74) Brumberg, A.; Harvey, S. M.; Philbin, J. P.; Diroll, B. T.; Lee, B.; Crooker, S. A.; Wasielewski, M. R.; Rabani, E.; Schaller, R. D. Determination of the In-Plane Exciton Radius in 2D CdSe Nanoplatelets *via* Magneto-Optical Spectroscopy. *ACS Nano* **2019**, *13*, 8589–8596.

(75) Zhang, Y.; Mascarenhas, A. Scaling of Exciton Binding Energy and Virial Theorem in Semiconductor Quantum Wells and Wires. *Phys. Rev. B - Condens. Matter Mater. Phys.* **1999**, *59*, 2040.

(76) Achtstein, A. W.; Schliwa, A.; Prudnikau, A.; Hardzei, M.; Artemyev, M. V.; Thomsen, C.; Woggon, U. Electronic Structure and Exciton-Phonon Interaction in Two-Dimensional Colloidal Cdse Nanosheets. *Nano Lett.* **2012**, *12*, 3151–3157.

(77) Benchamekh, R.; Gippius, N. A.; Even, J.; Nestoklon, M. O.; Jancu, J. M.; Ithurria, S.; Dubertret, B.; Efros, A. L.; Voisin, P. Tight-Binding Calculations of Image-Charge Effects in Colloidal Nanoscale Platelets of CdSe. *Phys. Rev. B - Condens. Matter Mater. Phys.* **2014**, *89*, 035307.

(78) Kelley, A. M.; Dai, Q.; Jiang, Z. J.; Baker, J. A.; Kelley, D. F. Resonance Raman Spectra of Wurtzite and Zincblende CdSe Nanocrystals. *Chem. Phys.* **2013**, *422*, 272–276.

(79) García-Santamaría, F.; Chen, Y.; Vela, J.; Schaller, R. D.; Hollingsworth, J. A.; Klimov, V. I. Suppressed Auger Recombination in "Giant" Nanocrystals Boosts Optical Gain Performance. *Nano Lett.* **2009**, *9*, 3482–3488.

(80) Hou, X.; Kang, J.; Qin, H.; Chen, X.; Ma, J.; Zhou, J.; Chen, L.; Wang, L.; Wang, L. W.; Peng, X. Engineering Auger Recombination in Colloidal Quantum Dots *via* Dielectric Screening. *Nat. Commun.* **2019**, *10*, 1750.

(81) Biadala, L.; Liu, F.; Tessier, M. D.; Yakovlev, D. R.; Dubertret, B.; Bayer, M. Recombination Dynamics of Band Edge Excitons in Quasi-Two-Dimensional CdSe Nanoplatelets. *Nano Lett.* **2014**, *14*, 1134–1139.

(82) Houtepen, A. J.; Hens, Z.; Owen, J. S.; Infante, I. On the Origin of Surface Traps in Colloidal II-VI Semiconductor Nanocrystals. *Chem. Mater.* **2017**, *29*, 752–761.

(83) Christodoulou, S.; Climente, J. I.; Planelles, J.; Brescia, R.; Prato, M.; Martín-Garciá, B.; Khan, A. H.; Moreels, I. Chloride-Induced Thickness Control in CdSe Nanoplatelets. *Nano Lett.* **2018**, *18*, 6248-6254.

(84) Luo, S.; Kazes, M.; Lin, H.; Oron, D. Strain-Induced Type II Band Alignment Control in CdSe Nanoplatelet/ZnS-Sensitized Solar Cells. *J. Phys. Chem. C* **2017**, *121*, 11136-11143.

(85) Cho, W.; Kim, S.; Coropceanu, I.; Srivastava, V.; Diroll, B. T.; Hazarika, A.; Fedin, I.; Galli, G.; Schaller, R. D.; Talapin, D. V. Direct Synthesis of Six-Monolayer (1.9 nm) Thick Zinc-Blende CdSe Nanoplatelets Emitting at 585 nm. *Chem. Mater.* **2018**, *30*, 6957–6960.

(86) Shornikova, E.; Biadala, L.; Yakovlev, D.; Sapega, V.; Kusrayev, Y.; Mitioglu, A.; Ballottin, M.; Christianen, P.; Belykh, V.; Kochiev, M.; Sibeldin, N. N.; Golovatenko, A. A.; Rodina, A. V.; Gippius, N. A.; Kuntzmann, A.; Jiang, Y.; Nasilowski, M.; Dubertret, B.; Bayer, M. Addressing the Exciton Fine Structure in Colloidal Nanocrystals: The Case of CdSe Nanoplatelets. *Nanoscale* **2018**, *10*, 646–656.

(87) Gao, J.; Zhang, J.; Van De Lagemaat, J.; Johnson, J. C.; Beard, M. C. Charge Generation in PbS Quantum Dot Solar Cells Characterized by Temperature-Dependent Steady-State

Photoluminescence. *ACS Nano* **2014**, *8*, 12814–12825.

(88) Zhang, J.; Tolentino, J.; Smith, E. R.; Zhang, J.; Beard, M. C.; Nozik, A. J.; Law, M.; Johnson, J. C. Carrier Transport in PbS and PbSe QD Films Measured by Photoluminescence Quenching. *J. Phys. Chem. C* **2014**, *118*, 16228–16235.

(89) Gilmore, R. H.; Liu, Y.; Shcherbakov-Wu, W.; Dahod, N. S.; Lee, E. M. Y.; Weidman, M. C.; Li, H.; Jean, J.; Bulović, V.; Willard, A. P.; Grossman, J. C.; Tisdale, W. A. Epitaxial Dimers and Auger-Assisted Detrapping in PbS Quantum Dot Solids. *Matter* **2019**, *1*, 250–265.

(90) Abdellah, M.; Karki, K. J.; Lenngren, N.; Zheng, K.; Pascher, T.; Yartsev, A.; Pullerits, T. Ultra Long-Lived Radiative Trap States in CdSe Quantum Dots. *J. Phys. Chem. C* **2014**, *118*, 21682-21686.

(91) Kushavah, D.; Mohapatra, P. K.; Singh, M.; Ghosh, P.; Vasa, P.; Rustagi, K. C.; Bahadur, D.; Singh, B. P. Exciton-Phonon Interaction and Role of Defect/Trap States in CdSe Quantum Dots. In *Materials Today: Proceedings*; **2016**, *3*, 3992-3996.

(92) Murphy, G. P.; Zhang, X.; Bradley, A. L. Temperature-Dependent Luminescent Decay Properties of CdTe Quantum Dot Monolayers: Impact of Concentration on Carrier Trapping. *J. Phys. Chem. C* **2016**, *120*, 26490-26497.

(93) Jones, M.; Lo, S. S.; Scholes, G. D. Quantitative Modeling of the Role of Surface Traps in CdSe/CdS/ZnS Nanocrystal Photoluminescence Decay Dynamics. *Proc. Natl. Acad. Sci. U. S. A.* **2009**, *106*, 3011-3016.

(94) Jing, P.; Zheng, J.; Ikezawa, M.; Liu, X.; Lv, S.; Kong, X.; Zhao, J.; Masumoto, Y. Temperature-Dependent Photoluminescence of CdSe-Core CdS/CdZnS/ZnS- Multishell Quantum Dots. *J. Phys. Chem. C* **2009**, *113*, 13545-13550.

(95) Bakulin, A. A.; Neutzner, S.; Bakker, H. J.; Ottaviani, L.; Barakel, D.; Chen, Z. Charge Trapping Dynamics in PbS Colloidal Quantum Dot Photovoltaic Devices. *ACS Nano* **2013**, *7*, 8771-8779.

(96) Van Der Stam, W.; Grimaldi, G.; Geuchies, J. J.; Gudjonsdottir, S.; Van Uffelen, P. T.; Van Overeem, M.; Brynjarsson, B.; Kirkwood, N.; Houtepen, A. J. Electrochemical Modulation of the Photophysics of Surface-Localized Trap States in Core/Shell/(Shell) Quantum Dot Films. *Chem. Mater.* **2019**, *31*, 8484-8493.

(97) Veamatahau, A.; Jiang, B.; Seifert, T.; Makuta, S.; Latham, K.; Kanehara, M.; Teranishi, T.; Tachibana, Y. Origin of Surface Trap States in CdS Quantum Dots: Relationship between Size Dependent Photoluminescence and Sulfur Vacancy Trap States. *Phys. Chem. Chem. Phys.*

**2015**, *17*, 2850-2858.

(98) Krause, M. M.; Kambhampati, P. Linking Surface Chemistry to Optical Properties of Semiconductor Nanocrystals. *Phys. Chem. Chem. Phys.* **2015**, *17*, 18882–18894.

(99) Mack, T. G.; Jethi, L.; Kambhampati, P. Temperature Dependence of Emission Line Widths from Semiconductor Nanocrystals Reveals Vibronic Contributions to Line Broadening Processes. *J. Phys. Chem. C* **2017**, *121*, 28537–28545.

(100) Mooney, J.; Krause, M. M.; Saari, J. I.; Kambhampati, P. Challenge to the Deep-Trap Model of the Surface in Semiconductor Nanocrystals. *Phys. Rev. B - Condens. Matter Mater. Phys.* **2013**, *87*, 081201.

(101) Jethi, L.; Mack, T. G.; Kambhampati, P. Extending Semiconductor Nanocrystals from the Quantum Dot Regime to the Molecular Cluster Regime. *J. Phys. Chem. C* **2017**, *121*, 26102-26107.

(102) Mooney, J.; Krause, M. M.; Kambhampati, P. Connecting the Dots: The Kinetics and Thermodynamics of Hot, Cold, and Surface-Trapped Excitons in Semiconductor Nanocrystals. *J. Phys. Chem. C* **2014**, *118*, 7730–7739.

(103) Delikanli, S.; Yu, G.; Yeltik, A.; Bose, S.; Erdem, T.; Yu, J.; Erdem, O.; Sharma, M.; Sharma, V. K.; Quliyeva, U.; Shendre, S.; Dang, C.; Zhang, D. H.; Sum, T. C.; Fan, W.; Demir, H. V. Ultrathin Highly Luminescent Two-Monolayer Colloidal CdSe Nanoplatelets. *Adv. Funct. Mater.* **2019**, *29*, 1901028.

(104) Marino, E.; Kodger, T. E.; Crisp, R. W.; Timmerman, D.; MacArthur, K. E.; Heggen, M.; Schall, P. Repairing Nanoparticle Surface Defects. *Angew. Chem., Int. Ed.* **2017**, *56*, 13795–13799.

(105) Rabouw, F. T.; Van Der Bok, J. C.; Spinicelli, P.; Mahler, B.; Nasilowski, M.; Pedetti, S.; Dubertret, B.; Vanmaekelbergh, D. Temporary Charge Carrier Separation Dominates the Photoluminescence Decay Dynamics of Colloidal CdSe Nanoplatelets. *Nano Lett.* **2016**, *16*, 2047–2053.

(106) Lin, C.; Gong, K.; Kelley, D. F.; Kelley, A. M. Size Dependent Exciton Phonon Coupling in CdSe Nanocrystals through Resonance Raman Excitation Profile Analysis. *J. Phys. Chem. C* **2015**, *119*, 7491–7498.

(107) Morello, G.; De Giorgi, M.; Kudera, S.; Manna, L.; Cingolani, R.; Anni, M. Temperature and Size Dependence of Nonradiative Relaxation and Exciton-Phonon Coupling in Colloidal CdTe Quantum Dots. *J. Phys. Chem. C* **2007**, *111*, 5846–5849.

(108) Bodunov, E. N.; Danilov, V. V.; Panfutova, A. S.; Simões Gamboa, A. L. Room-Temperature Luminescence Decay of Colloidal Semiconductor Quantum Dots: Nonexponentiality Revisited. *Ann. Phys.* **2016**, *528*, 272-277.

(109) Bodunov, E. N.; Simões Gamboa, A. L. Photoluminescence Decay of Colloidal Quantum Dots: Reversible Trapping and the Nature of the Relevant Trap States. *J. Phys. Chem. C* **2019**, *123*, 25515-25523.

(110) Bodunov, E. N.; Antonov, Y. A.; Simões Gamboa, A. L. On the Origin of Stretched Exponential (Kohlrausch) Relaxation Kinetics in the Room Temperature Luminescence Decay of Colloidal Quantum Dots. *J. Chem. Phys.* **2017**, *146*, 114102.

(111) Diroll, B. T.; Cho, W.; Coropceanu, I.; Harvey, S. M.; Brumberg, A.; Holtgrewe, N.; Crooker, S. A.; Wasielewski, M. R.; Prakapenka, V. B.; Talapin, D. V.; Schaller, R. D. Semiconductor Nanoplatelet Excimers. *Nano Lett.* **2018**, *18*, 6948-6953.

(112) Tessier, M. D.; Biadala, L.; Bouet, C.; Ithurria, S.; Abecassis, B.; Dubertret, B. Phonon Line Emission Revealed by Self-Assembly of Colloidal Nanoplatelets. *ACS Nano* **2013**, *7*, 3332-3340.

(113) Baikie, I. D.; Mackenzie, S.; Estrup, P. J. Z.; Meyer, J. A. Noise and the Kelvin Method. *Rev. Sci. Instrum.* **1991**, *62*, 1326.

(114) Zisman, W. A. A New Method of Measuring Contact Potential Differences in Metals. *Rev. Sci. Instrum.* **1932**, *3*, 367.

(115) Kelestemur, Y.; Guzelturk, B.; Erdem, O.; Olutas, M.; Erdem, T.; Usanmaz, C. F.; Gungor, K.; Demir, H. V. CdSe/CdSe$_{1-x}$Te$_x$ Core/Crown Heteronanoplatelets: Tuning the Excitonic Properties without Changing the Thickness. *J. Phys. Chem. C* **2017**, *121*, 4650–4658.

(116) Niehues, I.; Schmidt, R.; Drüppel, M.; Marauhn, P.; Christiansen, D.; Selig, M.; Berghäuser, G.; Wigger, D.; Schneider, R.; Braasch, L.; Koch, R.; Castellanos-Gomez, A.; Kuhn, T.; Knorr, A.; Malic, E.; Rohlfing, M.; Vasconcellos, S. F.; Bratschitsch, R. Strain Control of Exciton-Phonon Coupling in Atomically Thin Semiconductors. *Nano Lett.* **2018**, *18*, 1751–1757.

(117) Smith, A. M.; Mohs, A. M.; Nie, S. Tuning the Optical and Electronic Properties of Colloidal Nanocrystals by Lattice Strain. *Nat. Nanotechnol.* **2009**, *4*, 56–63.

(118) Naughton, M. S.; Kumar, V.; Bonita, Y.; Deshpande, K.; Kenis, P. J. A. High Temperature Continuous Flow Synthesis of CdSe/CdS/ZnS, CdS/ZnS, and CdSeS/ZnS Nanocrystals. *Nanoscale* **2015**, *7*, 15895–15903.

(119) Izquierdo, E.; Robin, A.; Keuleyan, S.; Lequeux, N.; Lhuillier, E.; Ithurria, S. Strongly Confined HgTe 2D Nanoplatelets as Narrow Near-Infrared Emitters. *J. Am. Chem. Soc.* **2016**, *138*, 10496–10501.

(120) Bernevig, B. A.; Hughes, T. L.; Zhang, S. C. Quantum Spin Hall Effect and Topological Phase Transition in HgTe Quantum Wells. *Science* **2006**, *314*, 1757-1761.

(121) Bae, W. K.; Lim, J.; Lee, D.; Park, M.; Lee, H.; Kwak, J.; Char, K.; Lee, C.; Lee, S. R/G/B/Natural White Light Thin Colloidal Quantum Dot-Based Light-Emitting Devices. *Adv. Mater.* **2014**, *26*, 6387–6393.

(122) Orosz, L.; Réveret, F.; Médard, F.; Disseix, P.; Leymarie, J.; Mihailovic, M.; Solnyshkov, D.; Malpuech, G.; Zuniga-Pérez, J.; Semond, F.; Leroux, M.; Bouchoule, S.; Lafosse, X.; Mexis, M.; Brimont, C.; Guillet, T. LO-Phonon-Assisted Polariton Lasing in a ZnO-Based Microcavity. *Phys. Rev. B* **2012**, *85*, 121201.

(123) Maragkou, M.; Grundy, A. J. D.; Ostatnick, T.; Lagoudakis, P. G. Longitudinal Optical Phonon Assisted Polariton Laser. *Appl. Phys. Lett.* **2010**, *97*, 111110.

(124) Dang, C.; Lee, J.; Breen, C.; Steckel, J. S.; Coe-Sullivan, S.; Nurmikko, A. Red, Green and Blue Lasing Enabled by Single-Exciton Gain in Colloidal Quantum Dot Films. *Nat. Nanotechnol.* **2012**, *7*, 335–339.

(125) de Mello, J. C.; Wittmannn, H. F.; Friend, R. H. An Improved Experimental Determination of External Photoluminescence Quantum Efficiency. *Adv. Mater.* **1997**, *9*, 230.

(126) Giannozzi, P.; Baroni, S.; Bonini, N.; Calandra, M.; Car, R.; Cavazzoni, C.; Ceresoli, D.; Chiarotti, G. L.; Cococcioni, M.; Dabo, I.; Dal Corso, A.; de Gironcoli, S.; Fabris, S.; Fratesi, G.; Gebauer, R.; Gerstmann, U.; Gougoussis, C.; Kokalj, A.; Lazzeri, M.; Martin-Samos, L.; *et al*. QUANTUM ESPRESSO: A Modular and Open-Source Software Project for Quantum Simulations of Materials. *J. Phys. Condens. Matter* **2009**, *21*, 395502.

(127) Giannozzi, P.; Andreussi, O.; Brumme, T.; Bunau, O.; Nardelli, Buongiorno, M.; Calandra, M.; Car, R.; Cavazzoni, C.; Ceresoli, D.; Cococcioni, M.; Colonna, N.; Carnimeo, I.; Dal Corso, A.; de Gironcoloi, S.; Delugas, P.; DiStasio Jr., R. A.; Ferretti, A.; Floris, A.; Fratesi, G.; Fugallo, G.; *et al.* Advanced Capabilities for Materials Modelling with Quantum ESPRESSO. *J. Phys. Condens. Matter* **2017**, *29*, 465901.

(128) Hamann, D. R. Optimized Norm-Conserving Vanderbilt Pseudopotentials. *Phys. Rev. B - Condens. Matter Mater. Phys.* **2013**, *88*, 1–10.

(129) Perdew, J. P.; Burke, K.; Ernzerhof, M. Generalized Gradient Approximation Made Simple.

*Phys. Rev. Lett.* **1996**, *77*, 3865–3868.

(130) Pack, J. D.; Monkhorst, H. J. Special Points for Brillouin-Zone Integrations. *Phys. Rev. B* **1976**, *13*, 5188.

For table of contents only

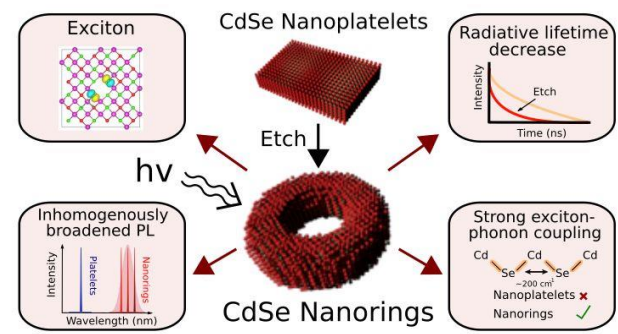